\documentclass[a4paper,fleqn]{cas-dc}

\usepackage[authoryear]{natbib}

\usepackage[english]{babel}
\usepackage{float}

\usepackage{multirow}
\usepackage{subcaption}

\def\tsc#1{\csdef{#1}{\textsc{\lowercase{#1}}\xspace}}
\tsc{WGM}
\tsc{QE}
\tsc{EP}
\tsc{PMS}
\tsc{BEC}
\tsc{DE}
\begin{document}
\let\WriteBookmarks\relax
\def\floatpagepagefraction{1}
\def\textpagefraction{.001}
\shorttitle{Constraining the Earth's Dynamical Ellipticity}
\shortauthors{Farhat et~al.}

\title [mode = title]{Constraining the Earth's Dynamical Ellipticity from Ice Age Dynamics }                      
%\tnotemark[1,2]

%\tnotetext[1]{This document is the results of the research
%   project funded by the National Science Foundation.}

%\tnotetext[2]{The second title footnote which is a longer text matter
%   to fill through the whole text width and overflow into
 %  another line in the footnotes area of the first page.}

\author{Mohammad Farhat}[type=editor,        orcid=0000-0001-7864-6627]
                      %  auid=000,bioid=1,
                     %   prefix=Sir,
                      %  role=Researcher,
                
\cormark[1]
%\fnmark[1]
%\ead{mohammad.farhat@obspm.fr}
%\ead[url]{www.cvr.cc, cvr@sayahna.org}

\address{IMCCE, CNRS, Observatoire de Paris, PSL University, Sorbonne Université, 77 Avenue Denfert-Rochereau, 75014, Paris, France}

\author{Jacques Laskar}

\author{Gwena\"{e}l Bou\'{e}}
%\fnmark[2]
%\ead{cvr3@sayahna.org}
%\ead[URL]{www.sayahna.org}

%\cortext[cor1]{Corresponding author}
%\cortext[cor2]{Principal corresponding author}
\cortext[1]{Email address: mohammad.farhat@obspm.fr}
%\fntext[fn2]{Another author footnote, this is a very long footnote and
%  it should be a really long footnote. But this footnote is not yet
%  sufficiently long enough to make two lines of footnote text.}

%\nonumnote{This note has no numbers. In this work we demonstrate $a_b$
%  the formation Y\_1 of a new type of polariton on the interface
%  between a cuprous oxide slab and a polystyrene micro-sphere placed
%  on the slab.
 % }

\begin{abstract}
 The dynamical ellipticity of a planet expresses the departure of its mass distribution from spherical symmetry. It enters as a parameter in the description of a planet's precession and nutation, as well as other rotational normal modes. In the case of the Earth, uncertainties in this quantity's history produce an uncertainty in the solutions for the past evolution of the Earth-Moon system. Constraining this history  has been a target of interdisciplinary efforts as it represents an astro-geodetic parameter whose variation is shaped by geophysical processes, and whose imprints can be found in the geological signal. We revisit the classical problem of its variation during ice ages, where glacial cycles exerted a varying surface loading that had altered the shape of the geoid. In the framework of glacial isostatic adjustment, and with the help of a recent paleoclimatic proxy of ice volume, we present the evolution of the dynamical ellipticity over the Cenozoic ice ages. We map out the problem in full generality identifying major sensitivities to surface loading and internal variations in parameter space. This constrained evolution is aimed to be used in future astronomical computations of the orbital and insolation quantities of the Earth.
\end{abstract}

%\begin{graphicalabstract}
%\includegraphics{figs/grabs.pdf}
%\end{graphicalabstract}

%\begin{highlights}
%\item Research highlights item 1
%%\item Research highlights item 3
%\end{highlights}


\maketitle

\section{Introduction}
The ellipsoidal flattening of the Earth, estimated to be around $1/299.627$ \citep{nakiboglu1982hydrostatic}, is a property of the mass distribution resulting from the hydrostatic competition between the dominant gravitational force and the weaker centrifugal force. Present observational inference of this quantity reports an excess in the flattening of $0.5\%$, corresponding to a difference between the equatorial and polar radii that is 100 m larger than equilibrium \citep{stacey2008physics}. This excess is also observed to be decreasing in an attempt of recovering  the equilibrium figure \citep{cox2002detection}. Mechanisms driving this excess vary in nature, magnitude, and time scales. They range from astronomical forcing leading to a variation in the gravitational potential  between the equator and the poles and tidal friction, to geophysical mechanisms pertaining to surface and internal adjustment in response to the mantle heterogeneity or to surface loading. The net outcome is altering the Earth's rotational motion and consequently the Milankovitch band cyclicity in proxy records \citep{levrard2003climate,stephenson2008historical, mitrovica2015reconciling}.

As a measure of the difference between the polar and equatorial moments of inertia, the dynamical ellipticity  is a global parameter that drives the precession and nutation  of the Earth. The difficulty in tracing its history over geological timescales is one of the major sources of uncertainty in determining the evolution of the Earth's obliquity and precession \citep{laskar1993orbital, laskar2004long}. In the last decades, proxy records corresponding to the recent millions of years have been used to attempt constraining the dynamical ellipticity variation \citep{palike2000constraints, lourens2001geological}.

We focus here on ice ages, during which cycles of glaciation and deglaciation exert a varying surface load upon the Earth's lithosphere due to the circulation of water between the ice caps and the oceans. This load forces the Earth to deform by subsiding under the growing load and rebounding upon its decay. However, this deformation is constrained by re-establishing an equilibrium between the crustal blocks and the underlying mantle; a state that is classically coined as "isostatic equilibrium". Thus glacial isostatic adjustment (GIA) describes the process of isostatic deformation due to ice and water surface loading variation. 

The influence of recent glacial cycles on the dynamical ellipticity have been first addressed outside the scope of GIA using simplified models of surface loading on a rigid Earth \citep{berger1988milankovitch,thomson1990quadratic,dehant1990potential}. The results are upper bound limits that cannot be attained in the context of a realistic Earth model, but were sufficient to the community at that time. More elaborate and meticulous approaches to the problem were motivated by the analysis of \cite{laskar1993orbital} suggesting a possible resonance capture with Jupiter and Saturn if the dynamical ellipticity is perturbed by a factor of -0.223\% relative to its present value. This called for a sequence of works that used climatic proxy records over the past million years to constrain the glacial surface loading, and the developed theory of the viscoelastic response of the Earth \citep{peltier1994procession, mitrovica1995pleistocene,mitrovica1997glaciation}. All studies concluded with the unlikelihood of occurrence of such an event. The problem was recently addressed again covering the past three million years corresponding to the interval of maximum glacial spread over the surface of the Earth \citep{morrow2012enigma,ghelichkhan2020precession}. 

In this work, we revisit the problem equipped with all the developments of the theory pertaining to the viscoelastic deformation of the Earth  and the self-consistent tracing of the glacial/oceanic surface loading in the context of the sea level equation formalism (Section \ref{GIA}) \citep{spada2019selen}. We use a very recent proxy record covering the Cenozoic era \citep{miller2020cenozoic}. This allows us to build a model of glacial loading variation over an extended interval of time (Section \ref{Ice_history}). Combining these two elements, we trace the variation of the perturbation in dynamical ellipticity by a set of numerical simulations of the sea level variation (Section \ref{delta_H}). We then map out the problem in full generality in an attempt to constrain its variables.

\section{Modelling Glacial Isostatic Adjustment}\label{GIA}
\begin{figure}
	\centering
		\includegraphics[width=0.4\textwidth, height=0.38\textwidth]{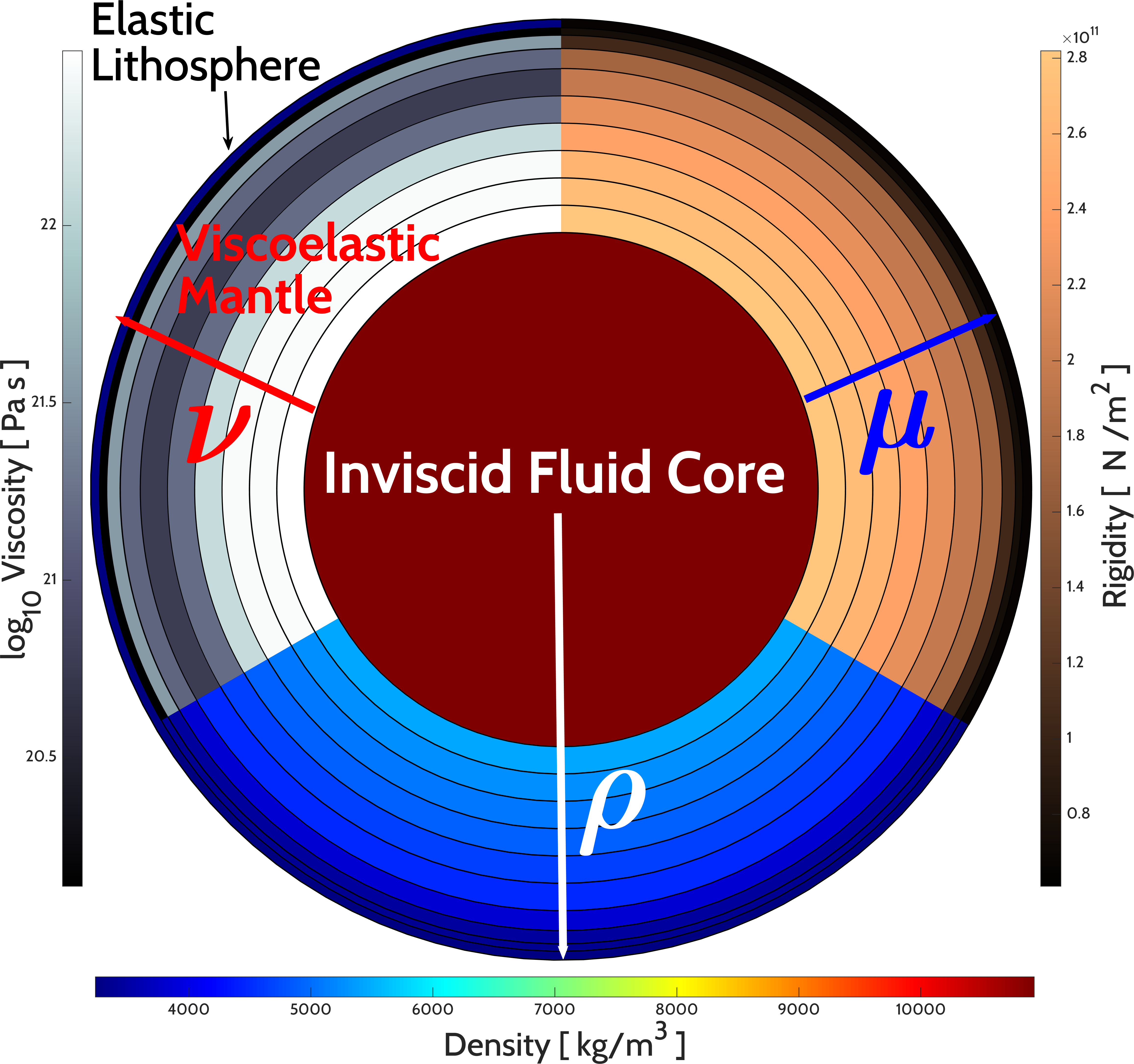}
	\caption{Internal radial stratification of the Earth discretized in an 11-layer model. Density and rigidity profiles are volume averaged from the PREM model \citep{dziewonski1981preliminary}, of which we adopted  the density discontinuities. The viscosity profile for the viscoelastic mantle is volume averaged from \cite{lau2016inferences}. The final layer corresponds to an elastic lithosphere of 90 km thickness. }
	\label{earth_interior}
\end{figure}

The governing, perhaps intuitive ingredients in modelling GIA are the spatio-temporal evolution of the surface loading and the rheological model of the Earth that dictates its response to that loading. We model a spherically symmetric Earth with radial physical properties based on the PREM model \citep{dziewonski1981preliminary}(Fig.\ref{earth_interior}), which behaves as an incompressible self-gravitating body. The core is modeled as a inviscid uniform fluid sheltered by the mantle that is approximated to behave viscoelastically \citep{wu1982viscous}. A choice has to be made on the rheology between linear (Newtonian), non-linear, or a transient power law describing the stress-strain relation \citep{gasperini2004linear}. We adopt the former description allowing the mantle to respond  as a linear Maxwellian body. Just below the surface, the lithosphere features  viscous deformation when forced over million of years \citep{england1986finite}, thus over glacial timescales ($	\lesssim 10^5$ years), it's safe and sufficient to assume that it behaves elastically. 

\subsection*{The viscoelastic response: Which internal stratification?}
When forced at the surface, the response of the model described above is quantified by the  Love numbers, which have elegant analytical forms when considering a homogeneous incompressible Earth (or the Kelvin Earth model \citep{munk1960rotation}). With the adequate stratified interior, closed form solutions are hard to obtain, thus the numerical normal modes theory has been adopted \citep{peltier1974impulse, Ver&Saba}. In this context, the Love numbers split into an elastic part that describes the instantaneous response to the impulsive load, and a viscous part  describing the relaxed response that is delayed in a multi-exponential form. We note that the elastic part is only dependent on the mass distribution and the rigidity of the body, while the viscous part is what carries the imprints of the chosen rheology. Thus the three surface loading Love numbers: $k^L, h^L, l^L$ associated with the gravitational perturbation, vertical, and horizontal displacements respectively take the general form
\begin{equation} \label{multi-exponentional}
    x_l^L(t) = x_l^{L,e} \delta (t) + H(t) \sum_{i=1}^N x_{l,i}^L e^{s_{l,i}t } 
\end{equation}
where $x_l^{L,e}$ is the elastic loading Love number of harmonic degree $l$, $H(t)$ is the Heaviside step function, $x^L_{l,i}$ are the viscoelastic residues associated with the $s_{l,i}$ normal modes in the framework of the viscoelastic normal mode theory. In the case where density only increases with depth and is invariant under deformation, these normal modes have real and negative values \citep{vermeersen1996analytical}, and they are associated with the viscous relaxation times at play, $\tau_{l,i}= -1/s_{l,i}$. The number of these normal modes $N$ is governed by the number of discontinuities (i.e. layer interfaces) in the radial profiles. Several numerical techniques and consequently several numerical codes are available for the computation of the Love numbers in the context of GIA, and a community benchmark study \citep{spada2011benchmark} tested the agreement of the results. We implemented the theory numerically using \textsc{Matlab}, and we solved the governing system of equations describing the Earth's deformation using the propagator matrix method (see \cite{sabadini2016global} for the full mathematical formalism). The obtained spheroidal deformation solution allows us to compute the loading Love numbers, while the normal modes are computed via solving the so-called secular determinant in the Laplace domain, making use of the correspondence principle \citep{peltier1974impulse}. We tested our code's precision against the benchmark study and we report an agreement of one part in $10^8$ for the Love numbers, and one part in $10^7$ for the normal modes.
 
\begin{figure}
    \centering
    \includegraphics[width=0.4\textwidth]{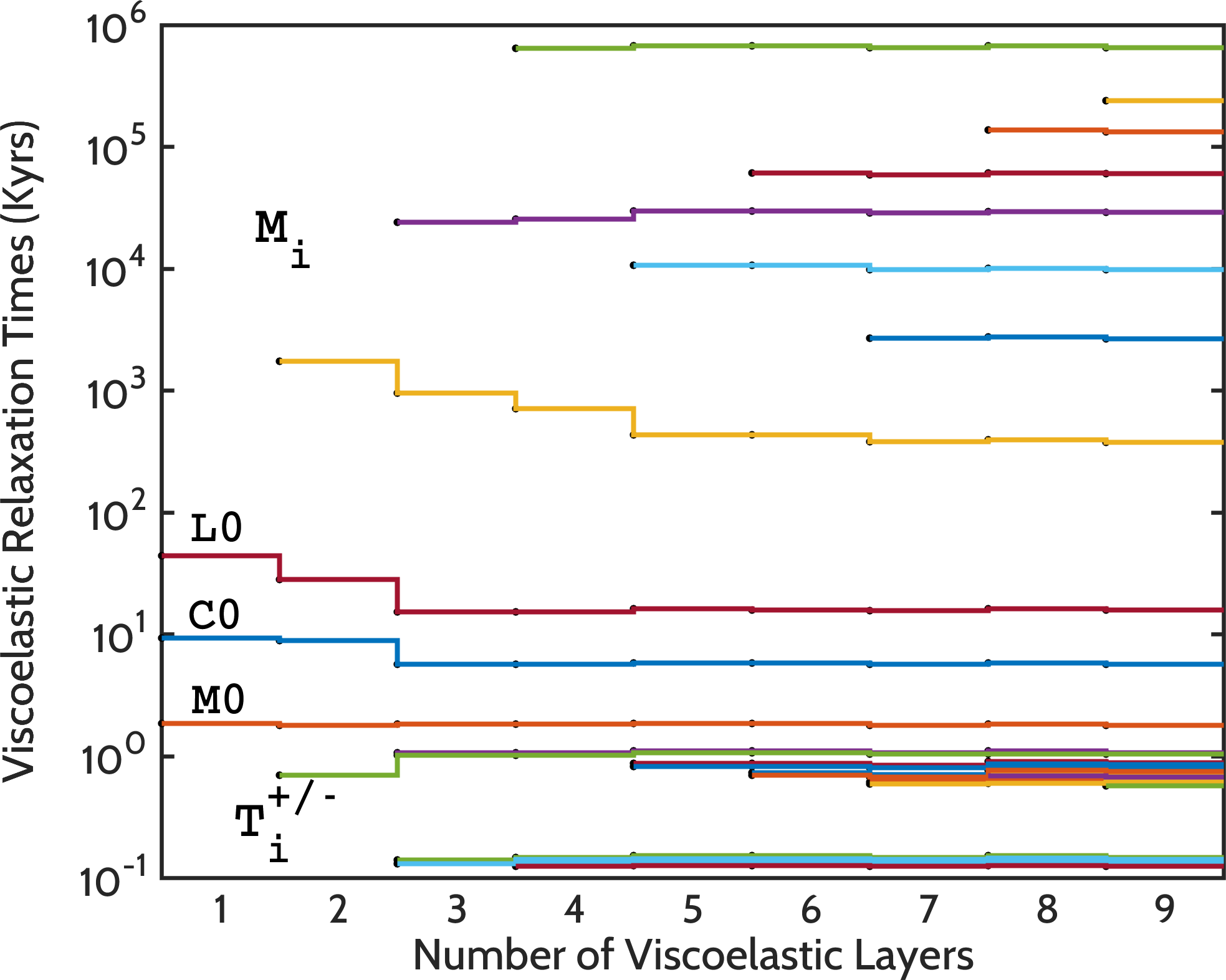}
    \caption{Viscoelastic relaxation times, $\tau_i= -1/s_i$, of harmonic degree $l=2$ computed in the normal modes theory formalism, for an increased stratification of the Earth's model. Zeroth modes are associated with the boundaries of the mantle with the core and the lithosphere. Buoyancy and transient modes, $M_i$ and $T_i$ doublets emerge at each added interface between viscoelastic layers, however their contribution to the multi-exponential form (Eq.\ref{multi-exponentional})  decays with increasing layering. The partitioning of the viscoelastic layers is consistent with the detected seismological discontinuities of PREM.  }
    \label{Relaxation_times_layers_fig}
\end{figure}

Using volumetric averages of the internal properties, a choice has to be made on the number of layers inside the Earth, specifically for the viscoelastic mantle  (Fig.\ref{earth_interior}). To better constrain this choice, we solve for the Love numbers and their associated relaxation spectra for an increasing level of stratification. In Fig.\ref{Relaxation_times_layers_fig} we summarize the variation of the relaxation times as a function of the number of viscoelastic layers. The radial viscosity profile was adopted from joint nonlinear inversions by \cite{lau2016inferences}, and volume averaged over the needed number of layers. As expected, the number of modes increases as we increase the number of layers (or analogously increasing the number of discontinuities). The $C0$, $L0$, and $M0$ modes, associated with the core-mantle and mantle-lithosphere boundaries, are of strongest viscous amplitudes and converge towards constant values for a number of layers as small as three. With each added viscoelastic interface, a buoyancy mode $M_i$ and a transient doublet $T_i^{\pm}$  emerge, though these modes feature  viscous amplitudes that decay as $i$ increases. A limitation of the normal modes theory lies in the arising difficulty when increasing the number of layers: the secular determinant computation and the roots finding algorithms might fail, which called for completely different approaches to solve for the Love numbers (e.g. \cite{spada2008alma}). But for our purposes here, we assume that a larger number of layers is not needed to capture the essential viscoelastic response of the Earth, as the modes of strongest amplitudes converge fast enough with stratification (see Fig.\ref{Relaxation_times_layers_fig}). Thus in our computations hereafter, we volume-average the viscoelastic mantle over nine layers, around the inviscid core, and covered by the elastic Lithosphere as shown in Fig.\ref{earth_interior}.

\subsection*{The Geoid's response and the sea level equation}
After computing the Love numbers, in order to fully characterize the rheological response to surface loading we use the Green's functions originally determined by \cite{peltier1974impulse, peltier1976glacial}. Specifically, the geoid Green's function takes the form \citep{wu1982viscous}
\begin{equation} \label{ggf}
    G^g (\theta,t) = \frac{a}{M_E} \sum_{l=0}^\infty \Big(\delta(t) +  k_l^L(t) \Big)P_l(\cos\theta)
\end{equation}
where $k_l^L$ is the gravitational loading Love number, $P_l$ is the set of fully normalized Legendre polynomials,  $\theta$ being the angular separation between the loading point and the response point, while $a$ and $M_E$ are the average radius and mass of the Earth. When describing the surface loading by a spatio-temporal variation function with respect to a reference state: $\mathcal{L}(\gamma,t)= L(\gamma,t) - L(\gamma,t_0)$, the geoid response function takes the general form
\begin{equation} \label{g srf}
    \mathcal{G} (\gamma,t) = G^g \otimes \mathcal{L}(\gamma,t)
\end{equation}
at a geographical position $\gamma= (\theta, \lambda)$, where $\otimes$ is a space-time convolution operator defined as
\begin{equation}\label{conv}
  G^g \otimes \mathcal{L}(\gamma,t)= \int_{-\infty}^{+\infty}dt^\prime \int_E   G^g(\alpha, t-t^\prime) \mathcal{L}(\gamma^\prime,t^\prime) dA^\prime
\end{equation}
where $\alpha$ is the angular separation between $\gamma$ and $\gamma^\prime$, $E$ is the Earth's surface, and $dA^\prime$ is an area element. We note that the calligraphic notation is used to indicate the variation of a quantity with respect to a reference state. Eqs.\eqref{g srf} and \eqref{conv} show that solving for the geoid variation requires a spatio-temporal evolution of the surface loading. The most accurate formalism used to track this evolution is that of the sea level equation (SLE), which traces the self-consistent gravitational redistribution of ice and water across the surface of the Earth. The SLE's mathematical theory and numerical implementation are undergoing continuous development by the community since its first introduction in the 1970s through a series of seminal papers \citep{peltier1974impulse,farrell1976postglacial, clark1978global, milne1998postglacial, mitrovica2003post, spada2007selen}, [see \cite{whitehouse2018glacial} for a clear historical review]. For completeness, we will only briefly describe the equation here from first principles. 
Given $\mathcal{N}$ as the variation of the height of the sea surface above the Earth's center of mass, and $\mathcal{U}$ as the vertical displacement of the Earth's solid surface, the spatio-temporal relative sea level variation is given by 
\begin{equation}
    \mathcal{S}(\gamma,t) = \mathcal{N}(\gamma,t)  - \mathcal{U}(\gamma,t) .
\end{equation}
Hence variations in the sea level are brought about by variations in the sea floor and the sea surface, which in turn respond to glacial and oceanic variations. This gives rise to the classical sea level response equation
 \begin{equation}\label{SLE}
        \mathcal{S} (\gamma,t) = \frac{\rho_i}{g} G^s \otimes_i \mathcal{I}(\gamma,t)   + \frac{\rho_o}{g}G^s \otimes_o  \mathcal{S}(\gamma,t) + C_{FC76} (t)
    \end{equation}
describing the variation of $\mathcal{S}$ relative to the continuously and viscoelastically deforming surface of the solid Earth;  where $\mathcal{I}(\gamma,t)$  is the variation in the spatio-temporal distribution of ice on the surface of the Earth; $\rho_i$ and $\rho_o$ are the densities of ice and water in the oceans; $G^s$ is a Green's function associated with the perturbation of the Earth's solid surface; $\otimes_i$ and $\otimes_o$ are convolutions in space and time over the ice sheets and oceans respectively, and $g$ is the surface gravity. Finally, $C_{FC76}(t)$, after \citep{farrell1976postglacial},  is a time dependent, but spatially invariant shift in sea level that is added to conserve the total mass transferred between the ice sheets and the oceans, and is given by
    \begin{equation}\label{CF76}
        C_{FC76} (t) = -\frac{m_i(t)}{\rho_o A^o(t)} -\frac{\rho_i}{g}\overline{G^s \otimes_i \mathcal{I}} 
     -\frac{\rho_o}{g}\overline{G^s \otimes_o \mathcal{S}}
    \end{equation}
where $A^o(t)$ is the varying ocean area and  $m_i(t)$ is the varying ice mass. The first term is often called the eustatic term, and it represents a surface average of sea level variation when the  four convolution terms in Eqs.\eqref{SLE} and \eqref{CF76} are dropped, or when the Green's functions vanish. This physically corresponds to the case of a rigid response and neglecting the time variations in the network of gravitational attractions between the solid earth, the oceans, and the ice sheets. The final two terms are spatial averages over the ocean, and are subtracted to ensure mass conservation. On the other hand, an extension to the equation was allowing for a dynamic  ocean area $A^o(t)$ by correcting for shorelines migration \citep{mitrovica2003post} and the extension of marine-grounded ice \citep{milne1998postglacial}.
    
The implicit nature of the SLE being a three dimensional non-linear integral equation,  with the sea level function present at both sides, taking the form of a Fredholm equation of the second kind, calls for an iterative approach in order to  be solved. We adopt the theory and numerical code developed over the past two decades and called \texttt{SELEN}$^4$ (Sea lEveL EquatioN solver, version 4.0) \citep{spada2007selen, spada2012modeling, spada2015selen, spada2019selen}. In \texttt{SELEN}$^4$, the SLE is solved by a pseudo-spectral iterative approach \citep{mitrovica1991postglacial} over a spatially discretized Earth surface on a spherical grid of icosahedron shaped pixels \citep{tegmark1996icosahedron}.  In the framework of the recent version of \texttt{SELEN}$^4$, the gravitationally self-consistent surface loading is computed allowing for shoreline migration and the transfer of ice between grounded and marine-based. For a more elaborate explanation on the \texttt{SELEN}$^4$ scheme, the reader is referred to the supplementary material of  \cite{spada2019selen}.    

Almost in all numerical implementations of the SLE, continuous solutions in time have not been found yet, and a time discretization of all the quantities is imposed. For example,  for the loading function we write
\begin{equation}
    \mathcal{L}(\gamma,t) = \sum_{n}^N \mathcal{L}_n(\gamma) H(t-t_n).
\end{equation}
With this time discretization and the pseudo-spectral approach, and using Eqs.\eqref{ggf} and \eqref{g srf}, the geoid response function, expanded in spherical harmonics of degree $l$ and order $m$, can then be written as
\begin{equation} \label{geoid_eqn}
      \mathcal{G}_{lm} =  \frac{3}{\rho^E} \sum_{n}^{N}  \frac{\Delta\mathcal{L}_{lm,n}}{2l+1} \Big( 1+k_l^{L,e} + \sum_{i=1}^{M} \frac{k_{l,i}^L}{s_{l,i}} (e^{s_{l,i}( t-t_n)} -1 )\Big) H(t-t_n) 
\end{equation}
with $ \Delta\mathcal{L}_{lm,n} = \mathcal{L}_{lm,n+1} - \mathcal{L}_{lm,n}$, and $\rho^E$ being the average density of the Earth. 
This geoid response can be decomposed into a rigid part
\begin{equation}
    \mathcal{G}_{lm}^r(t) = \frac{3}{\rho^E} \sum_{n}^{N}  \frac{\Delta\mathcal{L}_{lm,n}}{2l+1}H(t-t_n)
\end{equation}
which only accounts for elastic deformation when the elastic Love number is added, giving the form
\begin{equation}
     \mathcal{G}_{lm}^{e}(t) = \mathcal{G}_{lm}^r(t) ( 1+k_l^{L,e} ).
\end{equation}
Since the elastic Love numbers are between 0 and $-1$, this equation implies that the elastic deformation will always attenuate the loading effect on the rigid Earth by $|k_l^{L,e}|$ percent, to maintain the isostatic equilibrium. For example, for the second harmonic degree variation,  $k_2^{L,e}\approx-0.24$, meaning that the elastic deformation will compensate as much as 24\% of the effect on a rigid Earth. This will  further be augmented  by the delayed viscous relaxation through adding the viscoelastic residues. Taking the limit of infinite time after the surface loading is applied \citep{ricard1992isostatic}, the response function takes the form
\begin{align}\nonumber
         \mathcal{G}_{lm}^{ve}(t=\infty) &= \mathcal{G}_{lm}^r(t=\infty) \Big( 1+k_l^{L,e} -\sum_{i=1}^{M} \frac{k_{l,i}^L}{s_{l,i}} \Big)\\
                                    &= \mathcal{G}_{lm}^r(t=\infty) \Big( 1+k_l^{L,f} \Big)
\end{align}
where $k_l^{L,f}$ is the fluid loading Love number, which for the lowest degree harmonics is very close to negative unity. Specifically, $k_2^{L,f}\approx-0.98$, implying that the viscous relaxation can almost completely compensate for the surface loading after infinite time, where the remaining slight departure from perfect compensation is due to the presence of the elastic lithosphere \citep{wu1984pleistocene}.

Finding the variation in the geoid  allows us to compute the variation in the geopotential $\Phi$ using the classical Bruns formula \citep{heiskanen1967physical} 
\begin{equation}
   \mathcal{G} = \frac{\Phi}{g}.
\end{equation}
The multipolar expansion of the geopotential using the well known Stokes' coefficients \citep{yoder1995astrometric} allows us to write the geoid response function as 
\begin{equation}\label{geoid_stks_1}
    \mathcal{G}(\gamma,t) = a \sum_{l=2}^{l_{max}} \sum_{m=0}^{l} \big( \delta c_{lm}(t) \cos m\lambda + \delta s_{lm}(t)\sin m\lambda\big) P_{lm}(\cos\theta).
\end{equation}%
\begin{figure*}
	\centering
	  \includegraphics[width=\textwidth,height=3in]{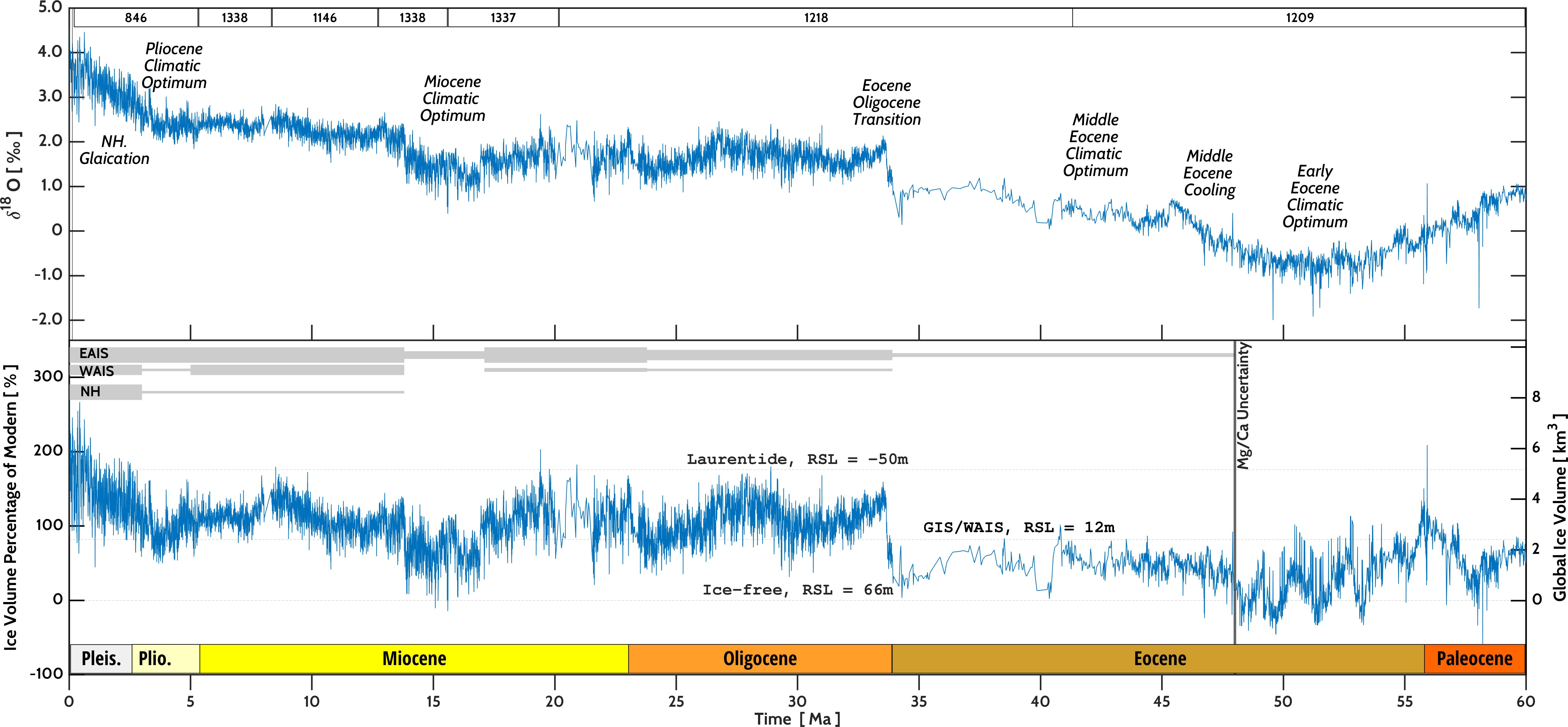}
	\caption{Cenozoic benthic foraminiferal $\delta^{18}O$ records from the splice compiled in \cite{miller2020cenozoic} (M20), and their ice volume contribution after the removal of the temperature contribution  à la \cite{cramer2011late}. The first panel is the isotope variation with time relative to the VPDB standard; the drilling sites are indicated on the top of the panel. The second panel is the ice variation with time in volume units on the right axis and in percentage relative to modern values on the left axis. The ice free line is equivalent to 66m of sea level increase above present, corresponding to the total increase in sea level if the Earth becomes ice free. The GIS/WAIS (Greenland and West Antarctic Ice Sheets) and the Laurentide lines are equivalent to 12m and -50m sea level variations. The gray sketches on top are rough estimates of the glacial evolution of the ice sheets.}
	    \label{O18_and_Ice}
\end{figure*}%
Thus after computing the harmonically decomposed geoid function over the discretized time history, the Stoke's coefficients can be obtained by using the general relationship between the coefficients of a complex spherical harmonics expansion of a time dependent scalar function, thus we have
\begin{equation}
    \delta c_{lm}(t) + \mathrm{ i} \delta s_{lm}(t) = \sqrt{2-\delta_{0m}} \mathcal{G}^*_{lm}(t)
\end{equation}
where the asterisk denotes complex conjugation. Using the scaling relationship derived by \cite{mitrovica1989pleistocene}, we find the gravitational zonal harmonics of the Earth as
\begin{equation}
    \delta J_l (t) = - \frac{1}{a} \sqrt{2l+1} \mathcal{G}_{l0} (t)
\end{equation}
and since we are after the harmonic corresponding to the equatorial flattening, we can write
\begin{equation}
    \delta J_2 (t) = -\frac{\sqrt{5}}{a} \delta c_{20}.
\end{equation}
The dynamical ellipticity of the Earth is a measure of the difference between the polar moment of inertia $C$ and the equatorial moments of inertia $A$ and $B$, namely
\begin{equation}
    H = \frac{C - (A +B)/2}{C}.
\end{equation}
It can also be written as  a linear function of the gravitational second zonal harmonic $J_2$
\begin{equation}
    H = \frac{M a^2}{C} J_2.
\end{equation}
Hence the variation of the dynamical ellipticity relative to its present day value $H_0\approx 3.27 \times 10^{-3}$ \citep{burvsa2008steady} can be written as
\begin{equation}
    \frac{\delta H (t)}{H_0} = \frac{\delta J_2(t)}{H_0 \mathcal{K}} = -\frac{\sqrt{5}}{H_0 \mathcal{K}a} \delta c_{20}
\end{equation}
where $\mathcal{K}= \frac{C}{Ma^2}$ is the so-called structure constant. Thus the problem of finding the relative variations in the dynamical ellipticity reduces for us to finding the geoid response function  $\mathcal{G}$ in the framework of the SLE solver. Another contribution arises from variations in the centrifugal potential, but we ignore that based on the arguments in Appendix \ref{app_A}.

\section{Cenozoic Ice History}
\label{Ice_history}
In addition to the Earth model, an ice loading history is required for the SLE solver to predict the evolution of the dynamical ellipticity with time. We use the most common  proxy of benthic foraminiferal measurements of oxygen isotopes ratio to constrain the global ice volume \citep{shackleton1975paleotemperature, zachos2008early}. The latter is known to have contributions from both ice volumes and water temperature, thus Mg/Ca benthic foraminiferal ratios have been recently used as an independent proxy for deep ocean temperature \citep{lear2000cenozoic, sosdian2009deep, cramer2011late}. The separation of the contributions can be established via a paleotemperature equation \citep{o1969oxygen, lynch1999geostrophic}. The ice contribution can then be scaled into a variation in global sea level using a certain calibration (see for example \citep{winnick2015oxygen}).  This calibration requires modelling as a function of the sizes of ice sheets \citep{raymo2018accuracy}, but it can serve as a first approximation to sea level variation. This sea level equivalence of ice can be scaled to a percentage of present ice volume.  Based on this technique, we adopt the oxygen isotope splice and its associated sea level equivalent of ice compiled recently in \citep{miller2020cenozoic}, covering the Cenozoic era, starting 66 Ma\footnote{We use the usual convention in stratigraphy: ka, Ma (thousand, million years) denote dates in the past from now, while kyr, Myr denote durations. }. This compiled splice, denoted M20 hereafter, is similar to that in \citep{de2017alternating}, but is composed entirely of Pacific records, minimizing the effects of temperature and salinity present in other regions due to deep circulation changes.

The M20 splice, Fig.\ref{O18_and_Ice}, can be interpreted to distinguish between a mostly unglaciated Cenozoic hothouse with $\delta^{18}O < -0.5$\textperthousand, a moderate greenhouse having ephemeral ice sheets with $-0.5< \delta^{18}O < 1.8$\textperthousand, and an ice house with continental scale ice sheets on one or both of the Earth's poles with higher isotopic values \citep{miller1987tertiary, huber2018rise}. A long term warming stage started in the late Paleocene (60 to 54 Ma) and led to the Paleocene-Eocene thermal maximum followed by a stable interval of minimum isotopic values during the Early Eocene climatic optimum (55 to 48 Ma). The middle Eocene after that witnessed a cooling phase with an increase of around 2\textperthousand, then the climate relatively stabilizes until the end of the Eocene with $\delta^{18}O \approx 1$\textperthousand. Thus during the Early Eocene, the Earth was largely ice-free, and high amplitude ice volume oscillations are most probably due to the error in the Mg/Ca record. This limitation is discussed in details in \citep{cramer2011late}, putting much larger errors on the record before 48 Ma, which can explain the negative ice volumes obtained before this period. This state was terminated by one of the major known Ma-scale features of the Cenozoic, the Eocene-Oligocene transition (EOT) around 34 Ma \citep{coxall2005rapid}, and is associated with the rapid glaciation of the Antarctic ice sheet (AIS) up to a continental scale,  marking the onset of the Earth's ice house.

\begin{table*}[width=1\textwidth,cols=2,pos=h] 
  \caption{A compilation of references  that were used to constrain the spatio-temporal distribution of ice over the Earth's surface. Some elements of the list correspond to geological evidence of various ice sheets extent over the Cenozoic. Others correspond to numerical modelling of ice sheets, specifically used to simulate onsets of continental glaciation. }
  \begin{tabular*}{\tblwidth}{@{}  LLLLLL@{} }
   \toprule
     Geologic Evidence / Geographic Constraint & Reference  \\
   \midrule
    \textbullet\ Partial glaciation in Antarctic high elevation regions during the Early Eocene.               & \cite{rose2013early}      \\
    \textbullet\ Antarctic expansion into marine terminating glaciers, specifically \\ around the Aurora subglacial basin.  & \cite{gulick2017initiation}      \\
    \textbullet\ Simulating the inception of the EAIS requires small ice caps on elevated plateaus.       & \cite{deconto2003rapid}       \\
    \textbullet\ Sediment rafting by glacial ice on the S-E end of Greenland dating back late Eocene. & \cite{eldrett2007continental}       \\
    \textbullet\ Middle Eocene episodic glaciation on Greenland from  ice-rafted Fe-oxide grains. & \cite{tripati2018evidence}  \\
    \textbullet\ Seismic stratigraphic evidence  for ice  in the Ross Sea during Oligocene-Miocene.             & \cite{bartek1992evidence} \cite{bart2003were}      \\
    \textbullet\ Oligocene grounded ice in the WAIS around and far from Marie Byrd Land.                      &  \citep{rocchi2006oligocene,sorlien2007oligocene}  \\
    \textbullet\ Terrestrial retreat of the EAIS during the Miocene Climatic Optimum.                                               & \cite{levy2016antarctic}       \\
    \textbullet\ Expansion of terrestrial ice across     the Ross Sea continental shelf around 24.5–24 Ma.      &  \cite{hauptvogel2017evidence}        \\
    \textbullet\ Terrestrial AIS stability for the past 8 Ma from cosmogenic isotope data. & \cite{shakun2018minimal} \\
    \textbullet\ Early Pliocene loss of ice from WAIS and Greenland.                                            &  \cite{naish2009obliquity}       \\
    \textbullet\ Substantial marine ice retreat in the EAIS  during Early Pliocene. & \cite{cook2013dynamic}       \\
    \textbullet\ Simulating the AIS evolution over the last 3 Myr: Separation between polar caps'\\  contributions to the global volume.
    &  \cite{pollard2009modelling}       \\
   \bottomrule
  \end{tabular*}
 
  \label{geologic_refs}
\end{table*}
The AIS was almost completely established on the Eastern terrestrial region (EAIS) \citep{galeotti2016antarctic}.  Its evolution during the Oligocene shows that it was not yet permanently developed, as we have large scale oscillations that peaked at the loss of more than 60\% of the AIS  after the EOT around 30 Ma, indicating its long term instability. The extent to which the West Antarctic ice sheet (WAIS) participated in the mostly unipolar Oligocene glaciation is largely unknown.

During the Early to mid-Miocene, major variations occurred in the Antarctic ice sheet volume and extent. The M20 splice estimates larger variations  than those proposed by \cite{pekar2006high} (50\% to 125\% of modern EAIS values).  General circulation models previously failed to completely simulate such large scale variations because of strong hysteresis effects and the glacial-interglacial symmetry. Moreover, after the continental scale ice spread is achieved, the resultant ice sheet is rather stable in the simulations. Modelling  this variability probably requires adding more atmospheric components to account for ice sheet-climate feedback \citep{gasson2016dynamic}. These large oscillations were punctuated by the Miocene Climatic Optimum (17 - 13.8 Ma). The latter was a period of reduced ice volume where near ice free conditions were attained around 15 Ma, probably establishing the most recent ice-free Earth.

Following this warm period, the Middle Miocene Climatic Transition (MMCT) involved three major steps of cooling and consequently sea level falls resulting in a permanent EAIS \citep{miller2020cenozoic} and a global ice volume higher that today ($\sim$120\%).  The ice volume then remained approximately constant until early Pliocene. This scenario of Antarctic stability for the past 8 million years is supported by cosmogenic isotope data from the Ross Sea \citep{shakun2018minimal}. During the last 3 Myr, blow-ups of ice volume were associated with sea level lowering around 110 m below present, indicating the onset of a continental scale northern hemispheric ice sheets. The largest of these blow-ups were during the past 800 kyr. It should be noted that the last glacial maximum (21$\sim $26 ka) is not only a local maximum of glaciation but a global one across the Cenozoic, associated with the maximum sea level drop ($\sim $130 m).

\section{Evolution of Dynamical Ellipticity}\label{delta_H}
 \begin{figure*}
	\centering
	  \includegraphics[width=\textwidth,height=3.7in]{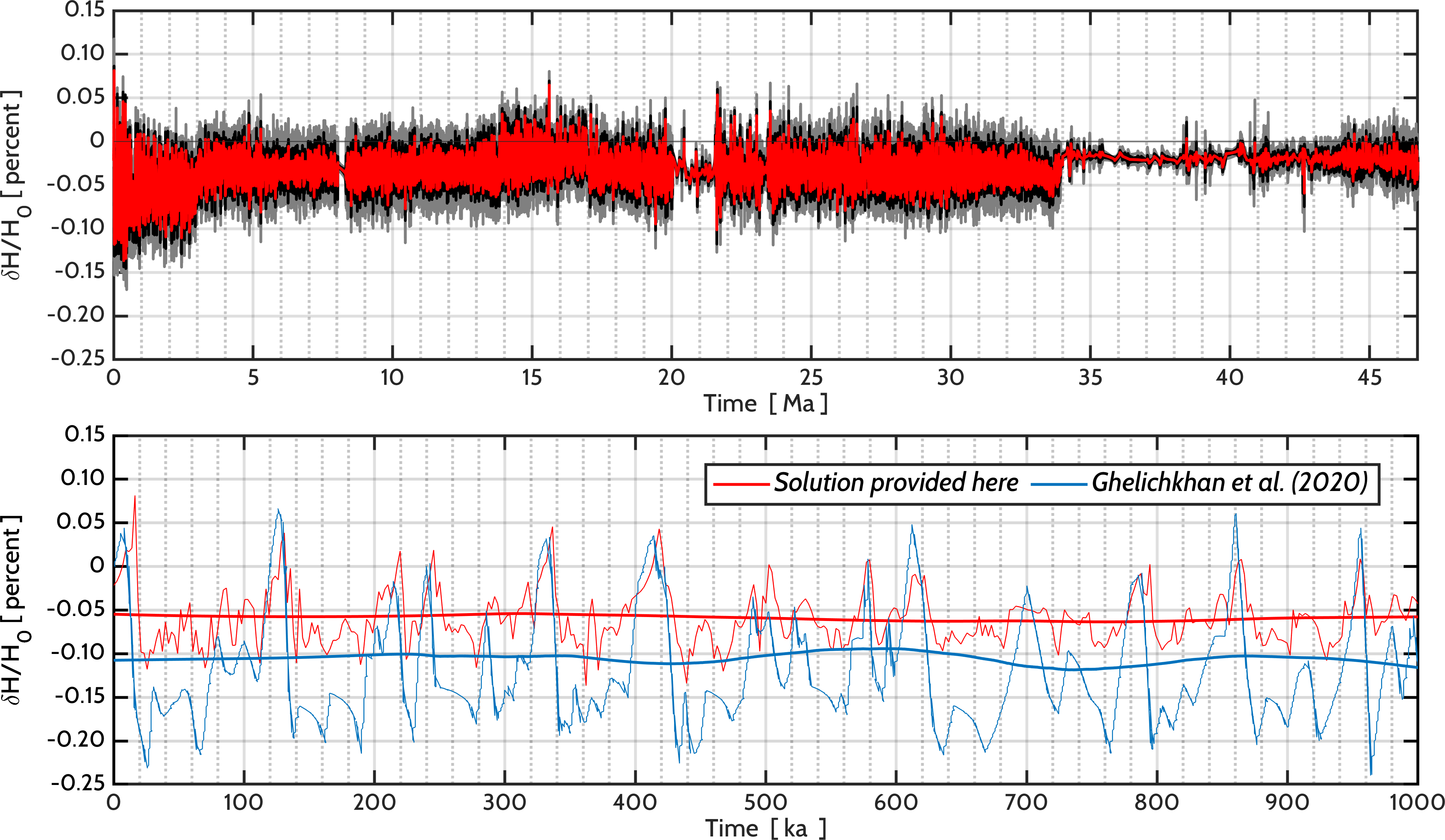}
	\caption{Evolution of the relative perturbation of dynamical ellipticty due to GIA over the past 47 Ma (the limit beyond which the error on Mg/Ca data is huge). \textit{Top:} Data in red corresponds to the ice input of Fig.\ref{O18_and_Ice} and viscosity profile of Fig.\ref{earth_interior} adopted from \cite{lau2016inferences} . In black and gray are 40 other simulations accounting for $1\sigma$ and $2\sigma$ uncertainty in ice as discussed in the text. \textit{Bottom:} A zoom over a smaller window over the past million years only. Plotted is our solution compared to that provided in \cite{ghelichkhan2020precession}, each with their smoothed secular trends.}
	    \label{dH_ice_ucertainty}
\end{figure*}
In the framework of the SLE solver, the ice volume input should be temporally discretized as we discussed, but also spatially distributed over the  surface of the Earth. Since an exact distribution is currently impossible to obtain over such a prolonged history, we approximate the input by conserving the global limit from the M20 splice (Fig.\ref{O18_and_Ice}), and abiding by major known glacial events and available geological constraints that can help to model the glacial spatial distribution. The latter are summarized in Table \ref{geologic_refs} (see Appendix \ref{app_B} for a more elaborate description  of the distribution). Such a distribution may not be adequate for high precision geodetic calculations. However, we are after the second degree harmonic decomposition of the load, which is characterized by even parity and is symmetric under rotation. Also, the change in oblateness reflects long wavelength deformation, so abiding by major climatic events, it is safe to assume that we would be capturing the backbone of the evolution of the dynamical ellipticity. Our sensitivity tests will later show that variations in the spatial distribution are only higher order corrections. 

Using \texttt{SELEN}$^4$, we discretize the Earth's surface onto a Tegmark  grid of equal-area icosaherdron-shaped pixels \citep{tegmark1996icosahedron}. The grid is characterized by a resolution parameter $R$ that yields a number of pixels $P=40R(R-1) +12$. In our suite of simulations, we set $R=30$, which gives  $P= 34812.$ We use the ICE-6G model \citep{argus2014antarctica} as a first approximation for the last glacial cycle spatial distribution, and then we scale it with time abiding by the global limit computed from the M20 splice (Fig.\ref{O18_and_Ice}) and the major climatic events (Table \ref{geologic_refs}). The SLE is solved over two nested loops, and the convergence of the solution as a function of the number of iterations is discussed in \citep{milne1998postglacial, spada2019selen}. Based on the convergence tests in these studies, all of  our simulations were performed over three internal and three external loops. We use the 11-Layers Earth's model described in Fig.\ref{earth_interior}, and we provide \texttt{SELEN}$^4$ with the needed Green's functions based on the viscoelastic response of this model. 

Global ice input uncertainty propagates from the uncertainty of the sea level variation. Estimates of uncertainty on the latter vary between $\pm 10$m and $\pm 20$m \citep{kominz2008late,miller2012high, raymo2018accuracy}. Thus we consider these limits as $1\sigma$ and $2\sigma$ error estimates, and we consequently create a white Gaussian noise with these amplitudes to perform 40 simulations of the SLE solver. In addition to this error, sea level variation is under a systematic uncertainty propagating from the variation in the volume of the ocean basin and a contribution from unconstrained tectonic changes. Constraining the former was done in \citep{cramer2011late} by limiting the ice contribution to the sea level variation from end member scenarios of complete ice melting \citep{lemke2007observations} and Airy loading \citep{pekar2002calibration}. To remove the tectonic contribution, we apply a LOESS regression filter \citep{cleveland1988locally} with a window of 20 Myr to keep only short timescale variability that is most likely due to ice volume variation.

In Fig.\ref{dH_ice_ucertainty}, we plot the evolution of the relative perturbation in dynamical ellipticity based on our SLE solutions. In black are solutions with $1\sigma$ correction, and $2\sigma$ solutions are in gray. Since the present Earth is in an interglacial period, the secular trend of the perturbation relative to the present day is a reduction in the flattening, as glaciation involves a net transfer of mass into the poles, reducing the flattening at the equator. As discussed earlier, the viscoelastic response attempts to compensate for this reduction by increasing the flattening again, but the overall perturbation nonetheless remains  negative.  This secular trend approaches zero during the Eocene, with relatively high amplitude oscillations  attributable to the poor constraint of the Mg/Ca ratios. 

The first major amplification in the perturbation occurs in a step-function like jump and is, as expected, across the Eocene-Oligocene transition (34 Ma), upon the initiation of a continental scale glaciation on Antarctica. After that, the unstable terrestrial East Antarctic Ice sheet results in moderate amplitude oscillations averaging around $-0.04\%$. The following major Ma-scale variation in the secular trend occurs around the Miocene climatic optimum when the Earth enters a period of reduced glaciation reaching near ice-free conditions. During this period, the relative perturbation in the ellipticity trend drops to around $-0.012\%$, then attains its global average again with the initiation of a larger scale glaciation on West Antarctica and Greenland, and with the stabilization of the EAIS. The final major variation in the trend occurs when the Earth transitions into its  bipolar glaciation. During the last 3 Myr, the dynamical ellipticity enters a  regime of extremely high amplitude oscillations that are maximized during the last million years. The secular trend during this period drops to around $-0.05\%$, and reaches $-0.07\%$ during the most recent glacial cycles. Glacial peaks over the same period average around $-0.11\%$, and reach $-0.17\%$ within the $2\sigma$ envelope. The last interglacial is marked with a global maximum  with a relative perturbation of $+0.1\%$. We note that using simpler geometries of glacial spread yields results that are well confined within this uncertainty envelope. For instance,  replacing the spatial evolution of the Antarctic glacial distribution by the spread of the LGM, which  almost represents a spherical cap confined within a circle of latitude at $-66^\circ$,  yields an evolution of the dynamical ellipticity within the $2\sigma$ uncertainty envelope for the Eocene and parts of the Miocene, and within $1\sigma$ for the rest of the Cenozoic.

In a similar procedure to that adopted here, \cite{ghelichkhan2020precession} also derived the evolution of the dynamical flattening due to GIA over the last 3 Myr. In the second panel of Fig.\ref{dH_ice_ucertainty}, we compare their solution to the present work over the past million years only. By visual inspection, the two solutions appear to evolve in-phase along the glacial cycles within the same order of magnitude. We also investigated the periodicity of both solutions and they matched identically. However, the evolution in \cite{ghelichkhan2020precession} involves more amplified oscillations and a larger secular reduction in the dynamical ellipticity. In fact, the plotted secular trends show that our estimate is around half that produced in their analysis (around $-0.055\%$ compared to $-0.11\%)$. Their study adopts an ice history from  \cite{raymo2011departures}, which is also developed from foraminiferal oxygen data. However, it is not clear whether their direct scaling took into account the contribution of temperature or not, so that could partially justify the discrepancy. However, as our ice sensitivity envelope well constrains the long term trend, we expect the difference to have emerged from adopting different viscosity profiles for the Earth, thus we perform a viscosity sensitivity analysis in the following sections. 

\subsection{Pacing by Astronomical Beats}
\begin{figure*}
	\centering
	  \includegraphics[width=\textwidth,height=3.2in]{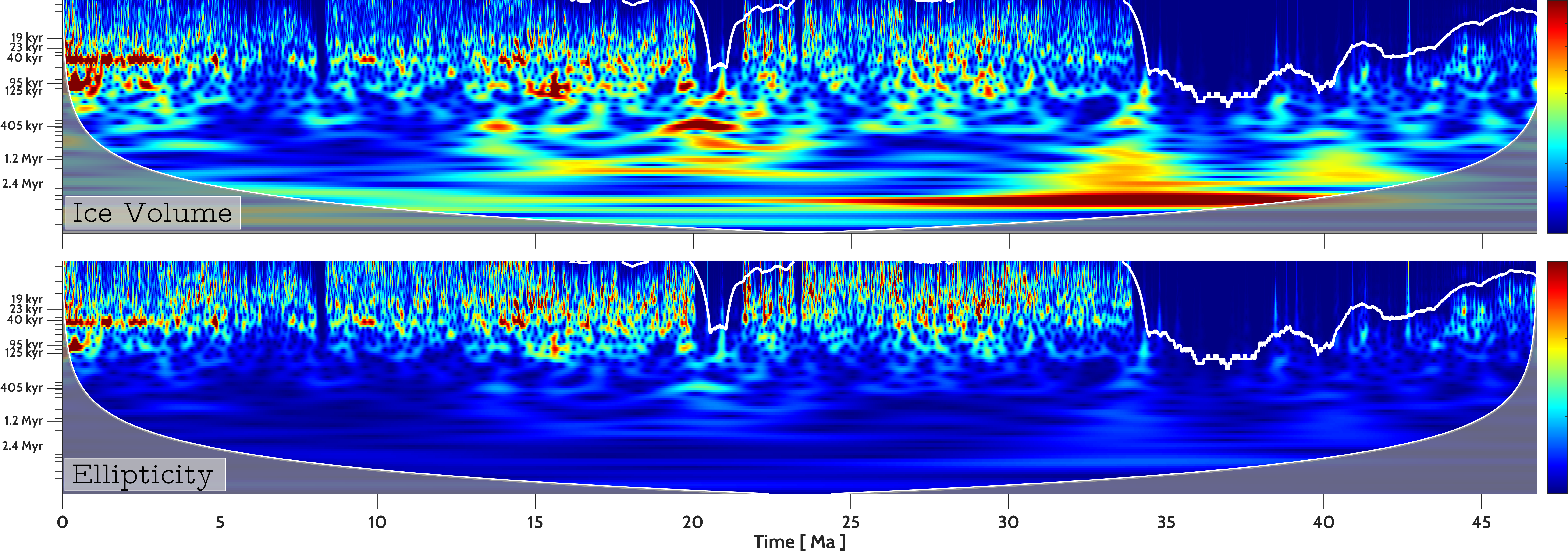}
	\caption{Continuous Wavelet Transforms (CWT) performed for the ice volume input provided to our simulations and the computed dynamical ellipticity evolution. The color mapping shows the relative power of varying amplitudes of spectral components of the data. Major spectral components associated with orbital forcing are identified on the y-axis. On top of the scalogram, the top white curve corresponds to the computed local nyquist frequency. The bottom shaded area represents the cone of influence, which is the area potentially affected by edge-effect artifacts, and is suspected to have time-frequency misinformation.  }
	    \label{cwt_IV_dH}
\end{figure*}

On time scales of $10^1\sim10^3$ kyr, the climatic state behaves as a nonlinear system that responds to quasi-periodic astronomical tuning. To better understand this modulation and its influence on the dynamical ellipticity variation, we perform a continuous wavelet transform (CWT) using \textsc{Matlab} for both the ice input data and our dynamical ellipticity evolution solution. That of the former is similar to that present in \citep{miller2020cenozoic}, and on a Myr timescale, it shows the general transition in the power spectrum from a climate that was mostly dominated by long period orbital forcing, into a regime of short period forcing dominance. The long eccentricity and obliquity cycles are clearly present before the Eocene-Oligocene transition, along with a less prominent shorter eccentricity ( $405$-kyr) modulation. The long periodicity dominance continued across the Oligocene, where large amplitude oscillations in ice volume were paced by the long obliquity cycle, along with an emerging dominance of the eccentricity period modulation \citep{boulila2011origin}. During this period, we also identify the short eccentricity ($100$-kyr) and obliquity (40-kyr) bands being present to a lesser power \citep{palike2006heartbeat, liebrand2017evolution}.  
\begin{figure}
    \centering
    \includegraphics[width=0.4\textwidth]{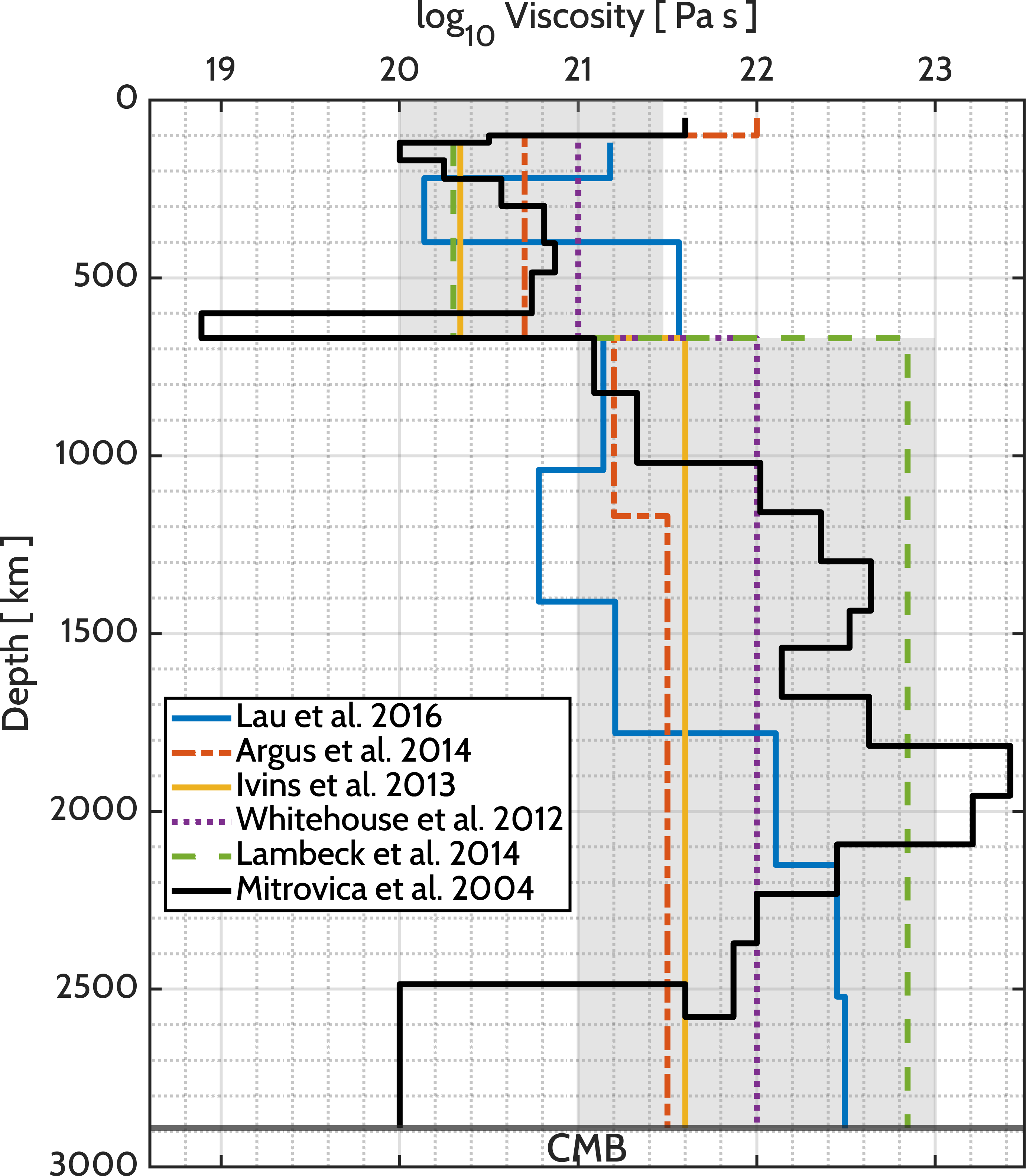}
    \caption{Mantle radial viscosity profiles from six different models in the literature. Most models are inferred from inversions of GIA observables including sea level variations, rebound, rate of change of the second zonal harmonic, and polar wander. All models involve at least an order of magnitude transition of viscosity between the mean of the upper mantle and the mean of the lower mantle, however some models advocate more acute jumps than others, especially around the 670 Km seismic discontinuity. The shaded areas are areas which we cover in our sensitivity analysis.   }
    \label{vis_profiles}
\end{figure}
\begin{figure}[h!]
    \centering
    \includegraphics[width=0.45\textwidth]{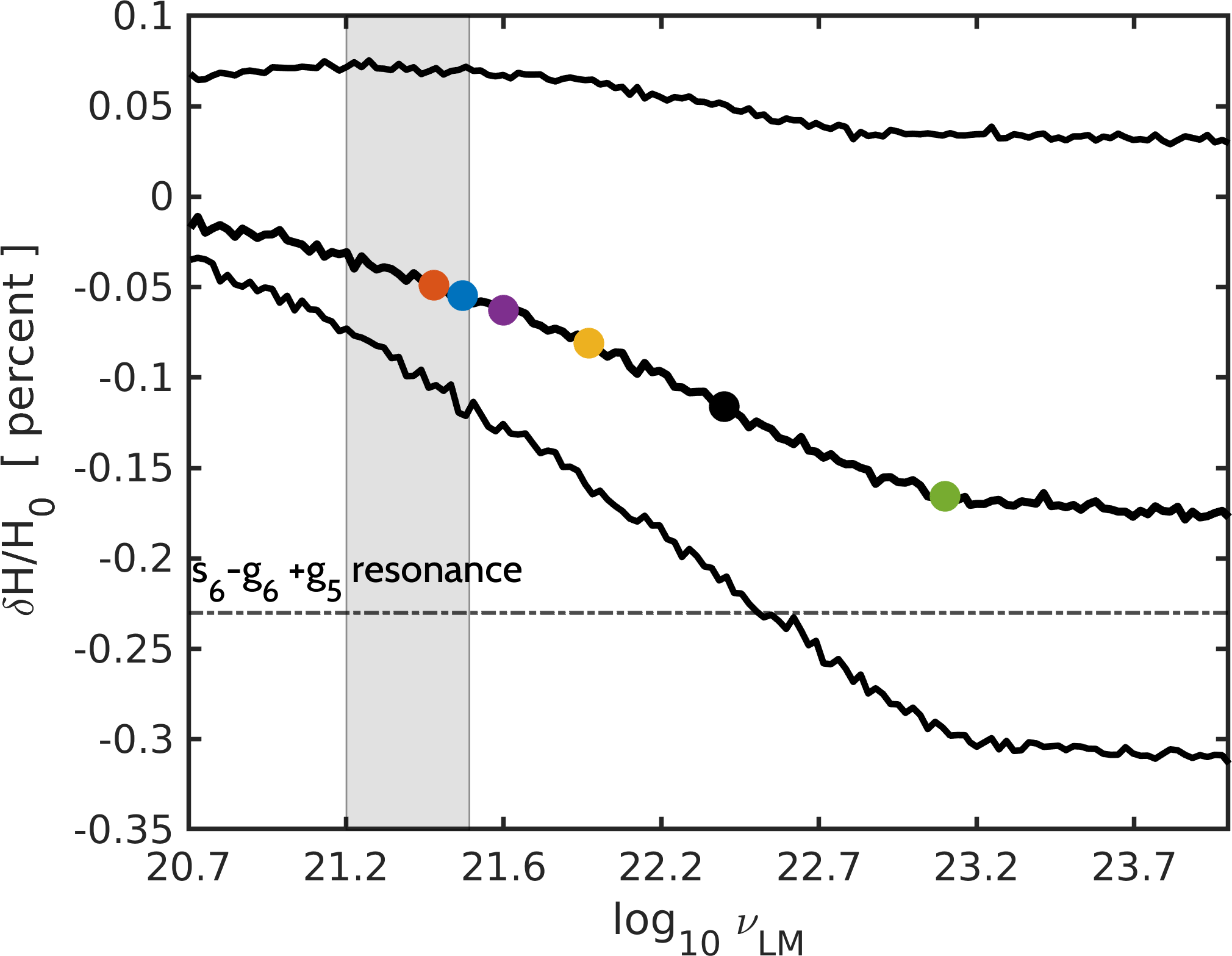}
    \caption{Relative perturbation in dynamical ellipticity over the past 3 Myr as a function of the lower mantle viscosity $\nu_{LM}$. The upper mantle viscosity is fixed at $\nu_{UM}=0.5\times10^{21}$ Pa~s. The middle branch represents the mean of the evolution, while the upper and lower branches represent the highest and lowest peaks in the evolution. For each value of $\nu_{LM}$ we perform 20 simulations that differ in the ice input to account for a random uncertainty in the interval [0 $2\sigma]$ as discussed in the text. Each point on the branches is thus the average of these simulations. Specified points on the branch refer to the specific viscosity profiles in Fig.\ref{vis_profiles} using the same color coding. The shaded area refers to a part of the identified region $\mathcal{R}_1$ after constraining the lower mantle viscosity to $\log_{10}\nu_{LM}\in [21.2, 21.5] $ by observed values of $\Dot{J}_2$  (see text).}
    \label{H_viscosity}
\end{figure}

Across the Miocene, the attenuation of the long period orbital forcing control is clear, in favor of a growing effect for the 405-kyr eccentricity and 40-kyr obliquity cycles. The emergence of the short obliquity forcing modulation is justified in \citep{levy2016antarctic} by the expansion of ice sheet margins into marine environments, which is a persistent feature after the Miocene Climatic Transition, 15 Ma. During the last 3 Myr, blow-ups of ice volume were associated with extreme sea level falls and the onset of a continental scale northern hemispheric ice sheets. The 40-kyr obliquity cycle continued to be dominant with the 100-kyr eccentricity cycle which takes over across the last 800 kyr, although it was already present before this transition. We note the clear attenuation of the 405-kyr eccentricity cycle that was dominant during intervals of the Miocene, and the almost complete muting of the long eccentricity and obliquity cycles. This general trend was also identified in another compiled oxygen splice \citep{westerhold2020astronomically}, explained by the spectrum of different nonlinear responses of the climate system to orbital forcing during different climate states: Eccentricity cycles should dominate the pacing of the Hothouse and the WarmHouse, as the eccentricity dominates temperature responses in low latitudes, while obliquity cycles dominate over the CoolHouse and the IceHouse, as high latitude glaciation is mostly influenced by the obliquity. This feature is also clear in our spectral analysis, except for the fact that long term obliquity pacing was also prominent even before the Oligocene. 

As for our dynamical ellipticity evolution, its CWT is a filtered version of that of the ice input. Since its evolution is dictated by the evolution of the surface loading, one can expect to have an identical pacing for both signals. However, the solid Earth's response behavior is orchestrated by the relaxation spectrum of the normal modes whose timescales range between $10^{-1}\sim10^5$ kyr (Fig.\ref{Relaxation_times_layers_fig}). We also note that the modes with the longest relaxation times correspond to buoyancy modes with very low normalized viscous amplitudes $k_{2,i}^L/s_{2,i}$, and thus minimal contribution to the summation in Eq.\ref{geoid_eqn}. Hence the viscous relaxation of the solid Earth acts as a high pass filter that will only keep short periodicities at play. Thus the CWT of the dynamical ellipticity attenuates the imprints of long orbital forcing, and maintains the pacing by the short obliquity and eccentricity cycles.
\subsection{Viscosity Profiles Sensitivity Test}

To better constrain the evolution of the dynamical ellipticity, we investigate the effect of mantle viscosity on the presented solution. The literature is very dense with modelled profiles, and we present a sample of them in Fig.\ref{vis_profiles}. The problem of inferring this radial profile from GIA observables dates back to \citep{daly1925pleistocene}. In general, relative sea level histories and post glacial rebound data, specifically those from Fennoscandia or Antractica, constrain the upper mantle's viscosity, while post glacial signals from Canada are used to constrain the upper part of the lower mantle. Other geophysical observables, including the rate of change of $J_2$ and the polar wander are used to constrain the viscosity of the rest of the lower mantle. Some radial profiles were derived from joint inversion of data that include these GIA effects along with data related to mantle convection \citep{mitrovica1997radial,mitrovica2004new, moucha2008dynamic}. Based on that, a community consensus has been established that the viscosity's radial profile increases some orders of magnitudes along the Earth's depth. However,  precise accounts of this transition are almost irreconcilable in the literature (Fig.\ref{vis_profiles}). Particularly, some models infer a viscosity jump of two orders of magnitude \citep{lambeck2014sea, nakada2015viscosity}, while others advocate a less acute transition (e.g. the VM5a model \citep{argus2014antarctica}). 

The viscosity profile we have used so far is presented in \cite{lau2016inferences}, and was constructed by analyzing GIA data using a combination of forward predictions and inversions based on nonlinear Bayesian inference. The result is constraining the upper mantle viscosity to around $3\times10^{20}$ Pa~s, the depth in between the mid-upper mantle and mid-lower mantle to around $10^{21}$ Pa~s, and the bottom half of the lower mantle to a mean value in excess of $10^{22}$ Pa~s. Rather than solving for the dynamical ellipticity for every model in the literature, we perform a systematic exploration of the solution's sensitivity to mantle viscosity variations. We perform simulations for  viscosity values within the shaded areas of Fig.\ref{vis_profiles} that encompass almost all of the profiles. 

As previously predicted by \citep{mitrovica1995pleistocene}, the results are mostly insensitive to viscosity variations in the upper mantle and almost merely dependent on the lower mantle. The contribution of the upper mantle is expected to arise for higher order harmonics. Thus in Fig.\ref{H_viscosity}, we show the results of a suite of simulations for a span of $\nu_{LM}$ for fixed $\nu_{UM}= 0.5 \times 10^{21}$ Pa~s. The simulations are trimmed over the last 3 Myr. We plot the mean, the maximum, and the minimum values of relative perturbations for each viscosity value. Each point on each branch is the average of 20 simulations that covered the ice uncertainty envelope. The overall trend of the mean variation monotonically decreases as we increase the viscosity jump value between the mantle parts. However, the slope of the decrease is smaller for smaller values of $\nu_{LM}$ than it is for larger values. The time evolution of each simulation also shows that the difference between the long term trends grows in time, justified by two reasons: the general feature of ice volume increase with time, and the accumulation of the non-linear effects due to the solid Earth's relaxation response. Besides the reduction in the mean, the figure shows that larger values of lower mantle viscosity, corresponding to a larger viscosity jump at the mantle parts' interface, amplify the response of the solid Earth to surface loading. This is seen in the broadening of the separation between the mean and the peaks.  This  increase in the reduction trend and the cycles' amplitude is understandable when increasing the viscosity. The latter results in an increase in the relaxation times which reduces the value of the fluid Love number, and consequently the magnitude of the viscoelastic compensation effect. The larger the viscosity value, the more we approach the limit of elastic compensation only, which is characterized by a larger relative perturbation in the geoid and the dynamical ellipticity. This justifies the plateau that we reach for  $\log_{10}\nu_{LM}>23.2$ Pa~s. On the contrary, decreasing the viscosity results in decreasing the relaxation times and increasing the viscoelastic compensation effect, thus shrinking the relative perturbation. Accounting for this viscosity effect can explain the discrepancy between the slightly larger amplitude oscillation in \cite{ghelichkhan2020precession}, as they use the viscosity profile based on the joint inversion of \cite{mitrovica2004new}, characterized by a larger lower mantle viscosity than our volume averaged profile from \citep{lau2016inferences}.

\begin{figure}
	\centering
	  \includegraphics[width=.5\textwidth]{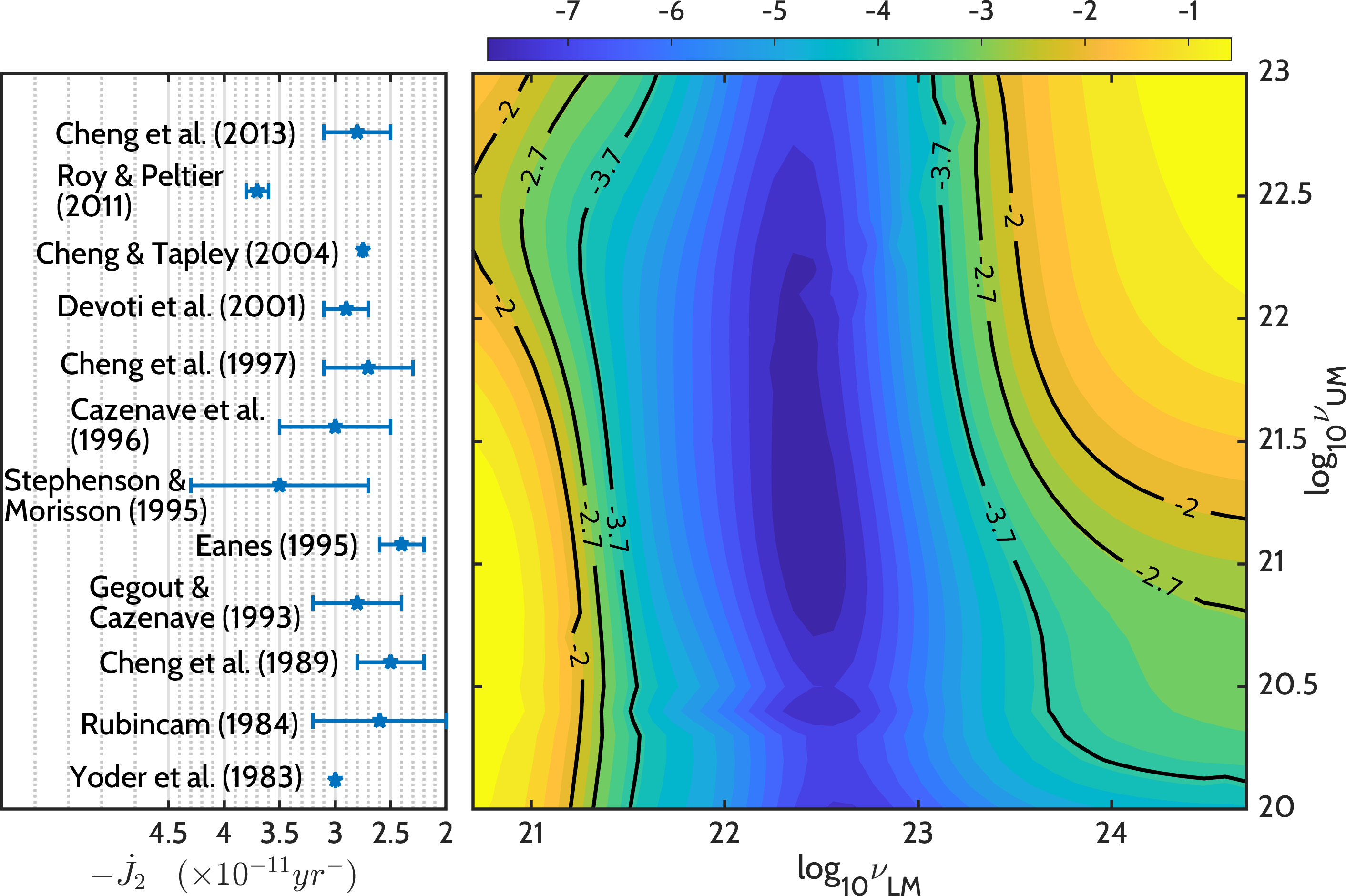}
	\caption{Constraining the choice of the viscosity profile by a comparison with observational estimates of $\Dot{J}_2$. Left: an inter-study of computed values of the secular trend. We note that more recent analyses were based on larger time spans of SLR measurements and more satellites. In studies that showed the quadratic form of the variation, we only took the negative linear trend that is most likely attributed to the post glacial rebound. We also note that, unlike the rest, the estimate in \cite{stephenson1995long} is based on the analysis of the Earth's rotational data over the recent centuries. Right: Level curves of the surface of computed $\Dot{J}_2$ on a grid of combinations of viscosity values for the upper and lower mantle parts. We specified some curves corresponding to relevant limits of observational values.}
	    \label{J2_inference}
\end{figure}

\subsection{Constraining by $\Dot{J}_2$ }
The average and the amplitude of the ellipticity perturbation cycles are highly sensitive to the chosen viscosity profile. The presented solution in Fig.\ref{dH_ice_ucertainty} is based on a recent inference of the viscosity profile. To justify this choice in the context of our viscosity sensitivity study, we attempt to constrain our viscosity freedom by the observed values of $\Dot{J}_2$, which as discussed earlier, is the major constraint used for lower mantle viscosity.  The first SLR-based estimate of the variation of $J_2$ was reported in \cite{yoder1983secular} as a linear trend of $-3\times10^{-11} yr^{-1}$. As time proceeded, further analysis of SLR data over longer time spans and using more satellites provided more estimates of the secular trend. In Fig.\ref{J2_inference} we compiled an inter-study comparison of this trend from different references. All studies until the late 1990s approximated the trend by a negative linear drift that is most likely an outcome of GIA. However, more recent analysis of the time span after 1995 showed a systematic decrease in this trend suggesting non-linearity \citep{cheng2013deceleration}. The likely cause of this swing was attributed to modern melting of glaciers as an outcome of global warming \citep{matsuo2013accelerated, loomis2019improved,chao2020variation}. Thus in the framework of GIA, we restrict our study to the  trends computed before the departure from linearity.

On a grid covering the ranges of mantle viscosity values (Fig.\ref{vis_profiles}), and using the already developed ice distribution, we compute present day rates of variation of $J_2$. In Fig.\ref{J2_inference}, we contour the surface of $\Dot{J}_2$ in this viscosity space, and we specify level curves of relevance with respect to the observed values. The latter appear to be concentrated around two regions, which as discussed earlier, are mostly dependent on the lower mantle viscosity. This sensitivity is clear with the vertical structure of the level curves.  Thus in total, two regions of viscosity combinations are preferred for the best fit with observational data: $\mathcal{R}_1$, a region enclosed by $\log_{10}\nu_{LM} \in [21, 21.5]$  for any value of $\nu_{UM}$, and a region $\mathcal{R}_2$, enclosed by $\log_{10}\nu_{LM}\in [23.1, 23.6]$ with large values of upper mantle viscosity. $\mathcal{R}_2$ diverges for lower values of upper mantle viscosity corresponding to an Earth with a very acute jump between the mantle parts (three to four orders of magnitude). Thus for values of upper mantle viscosity well constrained within the shaded region of Fig.\ref{vis_profiles}, $\mathcal{R}_1$ best fits the observed $\Dot{J}_2$. We identified $\mathcal{R}_1$ in the dynamical ellipticity evolution space in Fig.\ref{H_viscosity}. This analysis justifies our choice of the viscosity profile in the solutions presented earlier, though it was inferred from several geodetic parameter observations. A deviation from these well constrained regions arises when accounting for higher order harmonics comparison, and that can be justified in the presence of other mechanisms of surface and internal mass redistribution. 
\section{Summary and Conclusion}
In this paper, we provide the evolution of the dynamical ellipticity of the Earth over the past 47 Myr due to the varying glacial surface load. We revisit this problem identifying the major sensitivities of the dynamical ellipticity to surface loading and internal viscoelastic response. Concerning glacial history, we use a recently compiled far-field record of benthic oxygen isotopes that covers the Cenozoic \citep{miller2020cenozoic}. As both ocean temperature and ice volume take part in the isotopic variation, the contributions are deconvolved using benthic Mg/Ca records as an independent temperature proxy. We here consider that the glacial contribution of the isotopic record is an estimate of global ice volume, but we proceed with caution noting the following:
\begin{itemize}
    \item This is  only a rough estimate that gains more precision when accompanied by a record of oxygen isotope composition in the ice sheets \citep{langebroek2010simulating}.
    \item This estimate of ice history  could be compromised by variations in atmospheric moisture transport and the thickness of ice sheets yielding an overestimate of glacial volume \citep{winnick2015oxygen}.
    \item The comparison of the sea level equivalent of the record with the sea level construction from continental margins \citep{miller2020cenozoic} can also be compromised  by effects of mantle dynamic topography rather than pure glacial dynamics \citep{moucha2008dynamic}.
    \item However, our ice sensitivity tests proved that the correction to the dynamical ellipticity evolution due to an ice propagating error of $\pm 20$ m eustatic sea level equivalence does not produce a drastic change (Fig.\ref{dH_ice_ucertainty}). Moreover, errors arising from our transition from the far-field estimate to local distributions are minimized by the symmetries of the second zonal harmonic. These errors are  further minimized by the constraints we added from geological evidence (Table \ref{geologic_refs}).
\end{itemize}
Based on this ice history, we used the sea level equation solver of \cite{spada2019selen} to self-consistently trace the evolution of the surface loading between the ice caps and the oceans. 
On the other hand, the viscoelastic response of the Earth features the major variable in the problem. The evolution of the dynamical ellipticity mostly depends on the Earth's viscosity profile, specifically on the lower mantle viscosity. As the latter is not well determined in the literature, we studied this dependence thoroughly (Fig.\ref{H_viscosity}). We then proceeded by constraining the evolution in parameter space through a misfit analysis with recent observational estimates of $\Dot{J}_2$ (Fig.\ref{J2_inference}). The final outcome is constraining the average relative perturbation in the dynamical ellipticity over the past 3 Myr to $[ -0.031\%, -0.055\%]$,  with a maximal reduction inside $[ -0.07\%, -0.13\%]$, and a less sensitive  upper limit around $+0.07\%$. Trimming these estimates over the past 700 kyr, and using the same viscosity profiles, they fall in between the larger ellipticity reduction estimates of \cite{mitrovica1995pleistocene} and those less acute of \cite{peltier1994procession}. Our evolution extends to the  mostly unipolar interval of the Cenozoic with an average inside $[ -0.02\%$, $-0.045\%]$. Going beyond 47 Ma, our ice input, and consequently our ellipticity evolution are compromised by the growing error in the Mg/Ca record. 

This constrained history of the dynamical ellipticity will be used in the future long term numerical solutions for the orbital and rotational quantities of the Earth \citep{laskar2004long, laskar2011la2010}. Besides the surface loading, redistribution of mass within the Earth due to mantle convection also contributes to the evolution of the dynamical ellipticity. Such a contribution  also highly depends on the viscosity profile. However, different methods of recovering the mantle flow yielded vastly different results albeit using the same viscosity profile \citep{forte1997resonance, morrow2012enigma, ghelichkhan2020precession}. The surface loading effect alone precludes the possibility of a past capture into resonance with Jupiter and Saturn through the $s_6 - g_6  +g_5$ mode \citep{laskar1993orbital}. Thus the major uncertainty in the total variation of the dynamical ellipticity due to both effects leaves us with the need for more effort in this direction.

\section*{Acknowledgements}
This work was granted access to the HPC resources of MesoPSL financed
by the Region Ile de France and the project Equip@Meso (reference
ANR-10-EQPX-29-01) of the programme Investissements d’Avenir supervised
by the Agence Nationale pour la Recherche. It has also received funding from the European Research Council (ERC) under the European Union’s Horizon 2020 research and innovation programme (Advanced Grant AstroGeo-885250) and from the  Agence Nationale pour la Recherche (ANR) (Grant  AstroMeso ANR-19-CE31-0002-01).

\appendix
\section{Appendix: The Centrifugal Potential Variation} \label{app_A}
\renewcommand{\theequation}{A.\arabic{equation}}
\setcounter{equation}{0}
In addition to the direct effect of surface loading, the dynamical ellipticity is also subject to variation due to  the varying centrifugal potential, as a consequence of altering the magnitude and direction of the angular velocity vector. This rotational effect was one of the additions that came about during the development of the sea level variation theory. However, on the basis of the following approximations, we argue that this effect is minimal on the dynamical ellipticity, and we decided to ignore it in our calculations. 
Considering the rotating Earth and aligning the angular velocity vector $\vec\omega$ along the z-axis, one can write the centrifugal potential felt by a point P at the surface i.e. in the co-rotating frame as
\begin{equation}
    V^c (P) = \frac{1}{2}\omega^2 r_p^2 = \frac{1}{2}\omega^2 a^2 (1-\cos^2\theta_p)
\end{equation}
where $\theta_p$ is the angular separation of P from the axis of rotation. Using the Legendre polynomials, this can be written as
\begin{equation}
    V^c(P) = \frac{\omega^2 a^2}{3}\Big( P_0(\cos\theta_p) - P_2(\cos\theta_p)\Big).
\end{equation}
The theory usually continues by expanding the potential variation into variations in the angular velocity vector, and to leading terms in the perturbation,  this expansion is used in the problem of the true polar wander and solving Liouville equations \citep{sabadini2016global}. But for our purposes here, we are considering only the variations in the second zonal harmonic. In a general sense, the variation in the dynamical ellipticity $H$ due to variations in the  surface loading potential $V^L$, and centrifugal potential $V^c$ can be derived from
\begin{align}\nonumber \label{H_potnetials}
   H&= \frac{a^3}{GC}(1+k_2^L) V_2^L +\frac{a^3}{GC} k_2^T V_2^c\\
                &= \frac{a^3}{GC}(1+k_2^L) V_2^L  + k_2^T\frac{\omega^2a^5}{3GC}
\end{align}
where $a$ is the Earth's radius, $G$ is the gravitational constant, $C$ is the polar moment of inertia, $k_2^L$ and $k_2^T$ are the loading and tidal Love numbers, and $\omega=|\vec\omega|$.
%which when perturbed gives
%\begin{equation}
%    \delta H =\frac{a^3}{GC}(1+k_2^L) \delta V_2^L  + k_2^T\frac{\omega a^5}{3GC} \delta\omega
%\end{equation}
The variation in the potential is driven by variations in the polar moment of inertia on the following basis: 
A rigid Earth variation written as
\begin{equation}
    \delta C^r= \frac{2a^3}{3G}\delta V_2^L
\end{equation}
the viscoelastic compensation depicted in
\begin{equation}
    \delta C^{ve} =\frac{2a^3}{3G} k_2^L \delta V_2^L =  k_2^L  \delta C^r
\end{equation}
and the rotation effect 
\begin{equation}
    \delta C^{rot} = \frac{2a^3}{3G} k_2^T \delta V_2^c = \frac{4a^5}{9G} k_2^T \omega \delta\omega.
\end{equation}

In a general sense too, the dynamical ellipticity can be split into the hydrostatic component $H^f$, and a residual part $\delta H$ depending on all possible surface and internal irregularities
\begin{equation}
    H= H^{f}  +\delta H
\end{equation}
where the hydrostatic part can be found by taking the fluid limit of Eq.\eqref{H_potnetials}
\begin{align}\nonumber
    H^f =& \frac{a^3}{GC}(1+k_2^{L,f} ) V_2^L  + k_2^{T,f} \frac{a^5}{3GC}\omega^2 \\ 
    \approx& k_2^{T,f}\frac{a^5}{3GC}\omega^2 
\end{align}
where the last approximation is possible since $k_2^{L,f}\approx-0.98$ while $k_2^{T,f}\approx0.97$. Ignoring the loss of angular momentum due to tidal dissipation, the variation in the angular velocity or the length of the day is accompanied by a variation in the moment of inertia to conserve the angular momentum, thus we can write
\begin{align} \nonumber
\delta (C\omega) &= C\delta\omega + \omega(1+k_2^L)  \delta C^r + \omega\delta C^{rot} \\\nonumber
                &=\delta\omega\Big( C + \frac{4a^5}{9G} k_2^T \omega^2\Big)  + (1+k_2^L)\omega \delta C^r \\\nonumber
                &=\delta\omega C \Bigg(  1 + \frac{4}{3}\frac{k_2^T}{k_2^{T,f}}H^f \Bigg) + (1+k_2^L)\omega \delta C^r \\
                &\approx \delta\omega C + (1+k_2^L)\omega \delta C^r \equiv 0
\end{align}
where the approximation was based on the relatively small value of $H^f$. This allows us to write
\begin{equation}
    \delta\omega\approx - (1+k_2^L)\omega \frac{\delta C^r}{C}.
\end{equation}
With these quantities, we can finally compute the ratio of the contributions to the polar inertia variation  as
\begin{equation}
    \Bigg|\frac{\delta C^{rot}}{(1+k_2^L)\delta C^r}\Bigg| = \Bigg|\frac{4a^5\omega^2}{9GC} k_2^T\Bigg| = \Bigg|\frac{4k_2^T}{3k_2^{T,f}}H^f\Bigg| \approx 10^{-3}
\end{equation}
where $k_2^T\approx 0.3$, $k_2^{T,f}\approx0.97$, and $H^f\approx 3\times 10^{-3}.$ On this basis, we ignored the effect of rotation on the perturbation of dynamical ellipticity in our simulations. 

\section{Appendix: The Cenozoic Spatial Distribution of Ice} \label{app_B}
\renewcommand{\theequation}{B.\arabic{equation}}
\setcounter{equation}{0}
\begin{figure}[ht!]
\centering
        \includegraphics[width=0.3\textwidth, height=4.25cm]{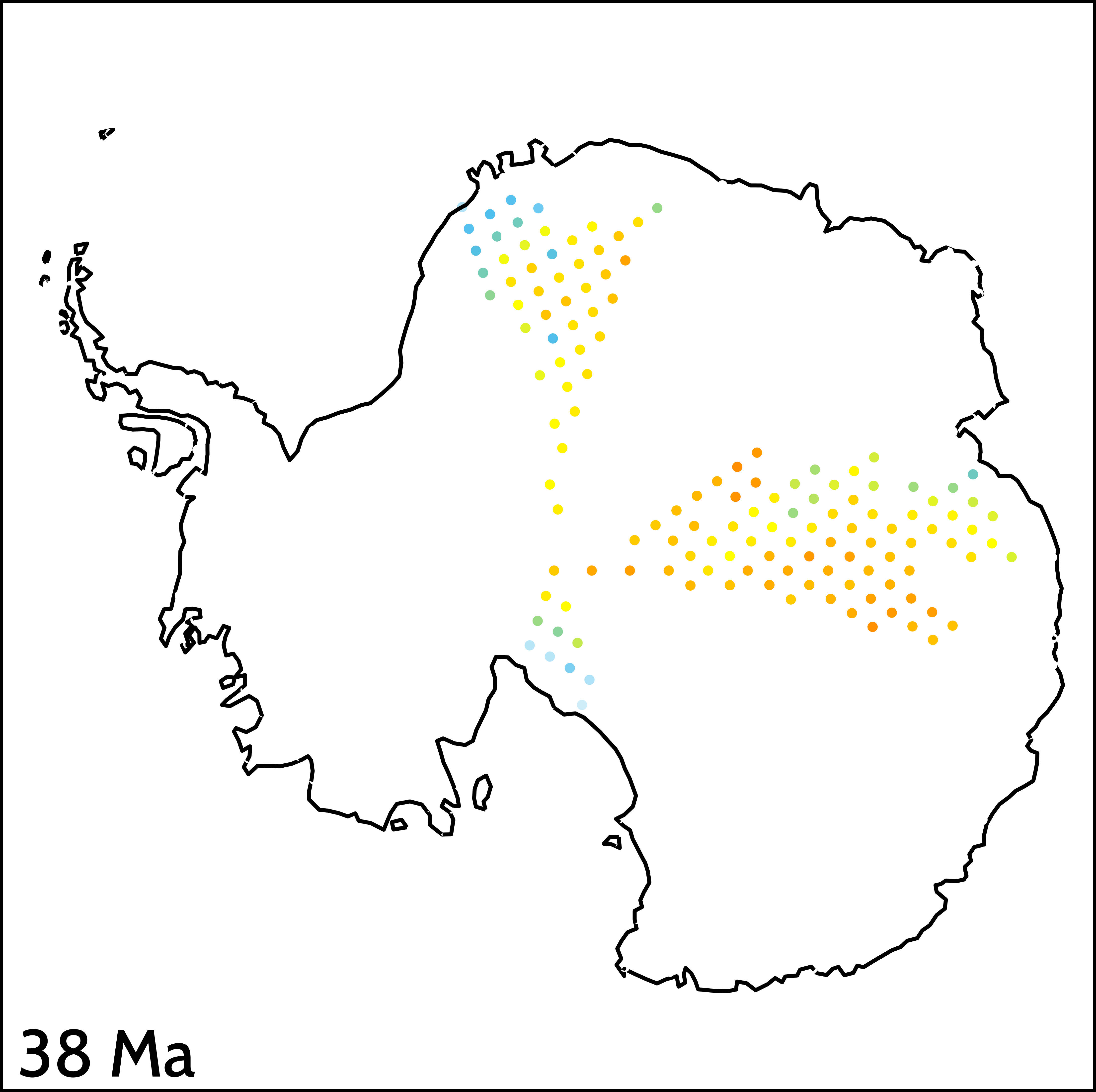}\hfill\\
           \includegraphics[width=0.3\textwidth, height=4.25cm]{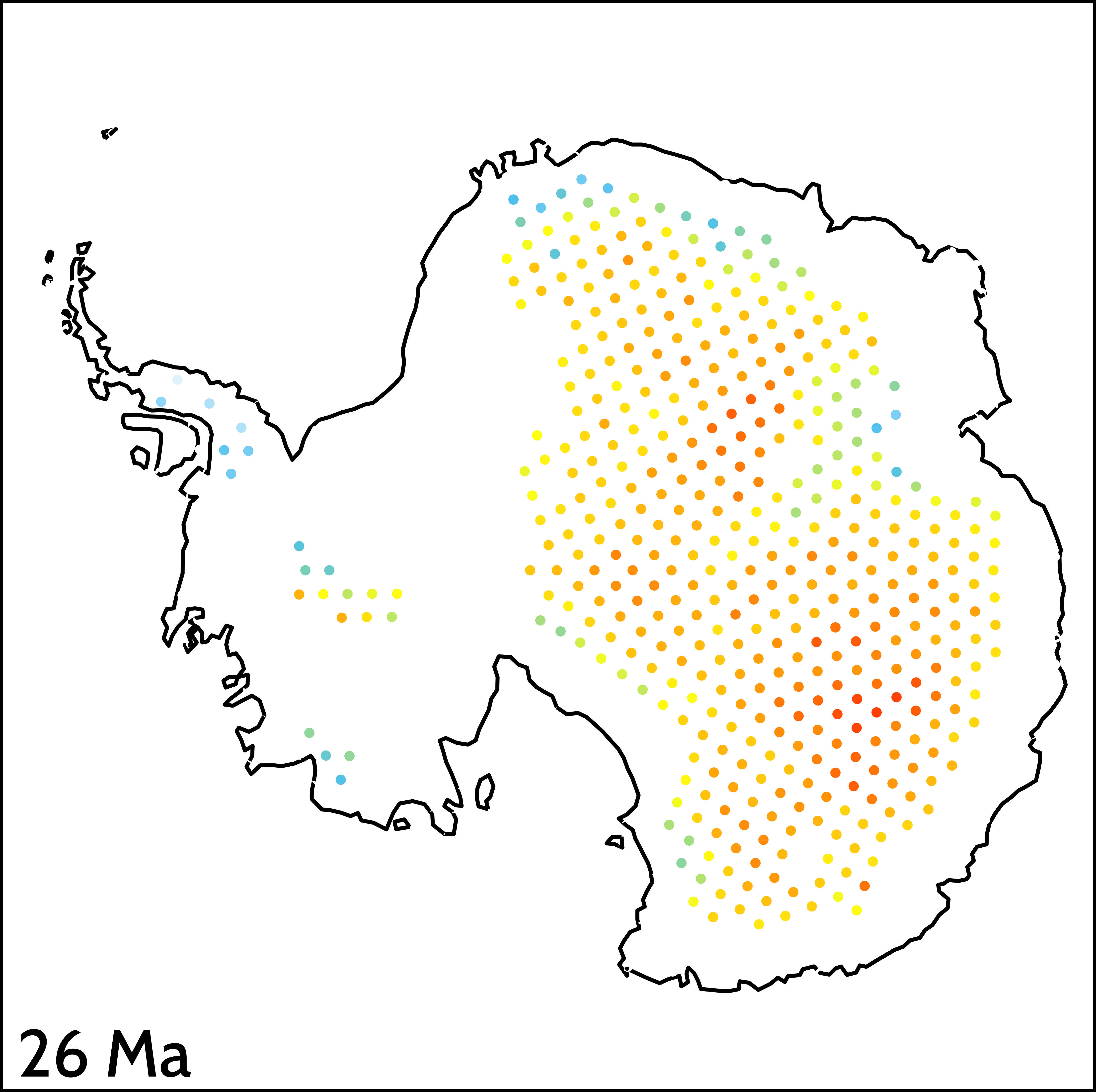}\hfill\\
              \includegraphics[width=0.3\textwidth, height=4.25cm]{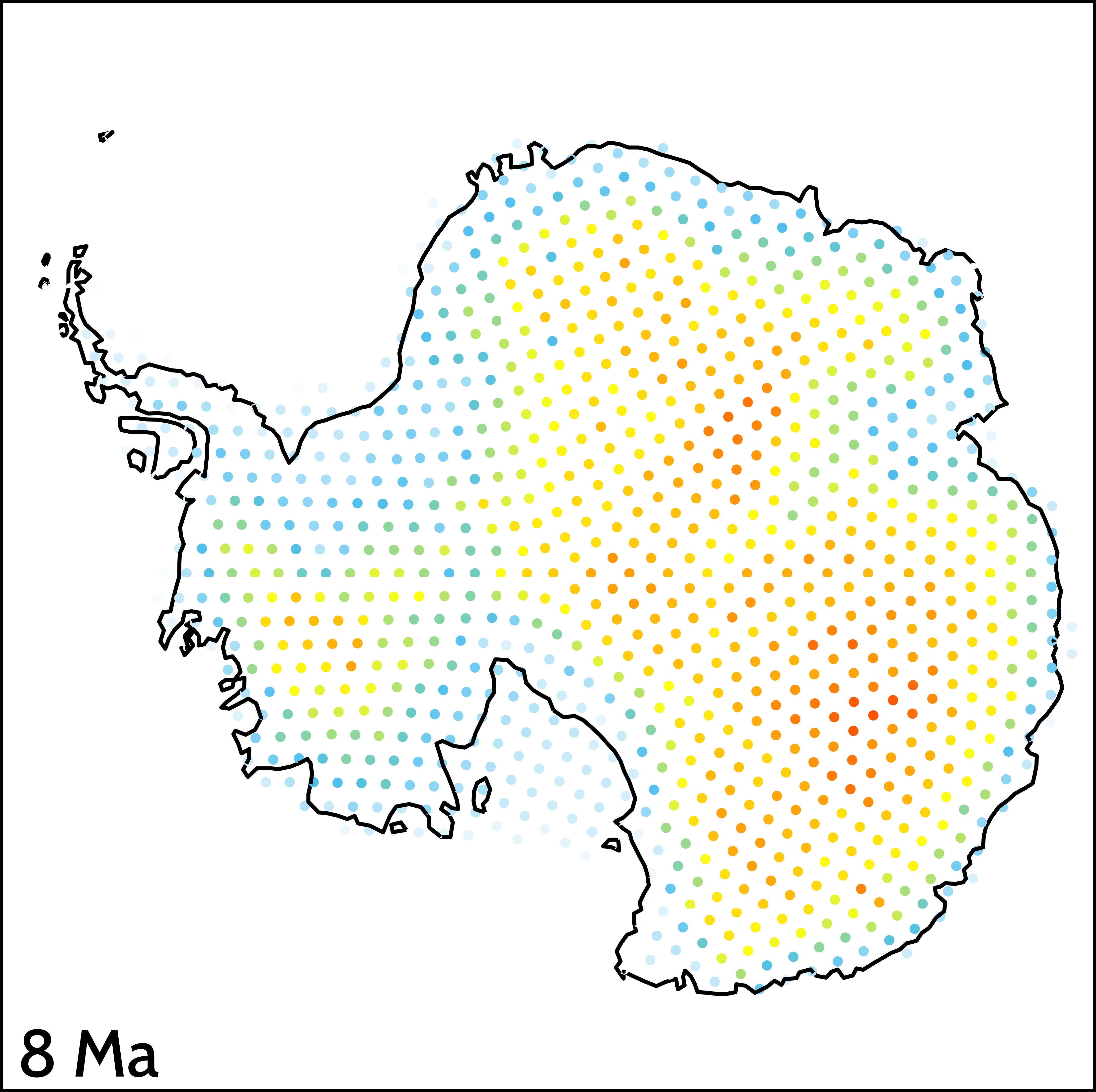}\hfill\\
            \includegraphics[width=0.3\textwidth, height=4.25cm]{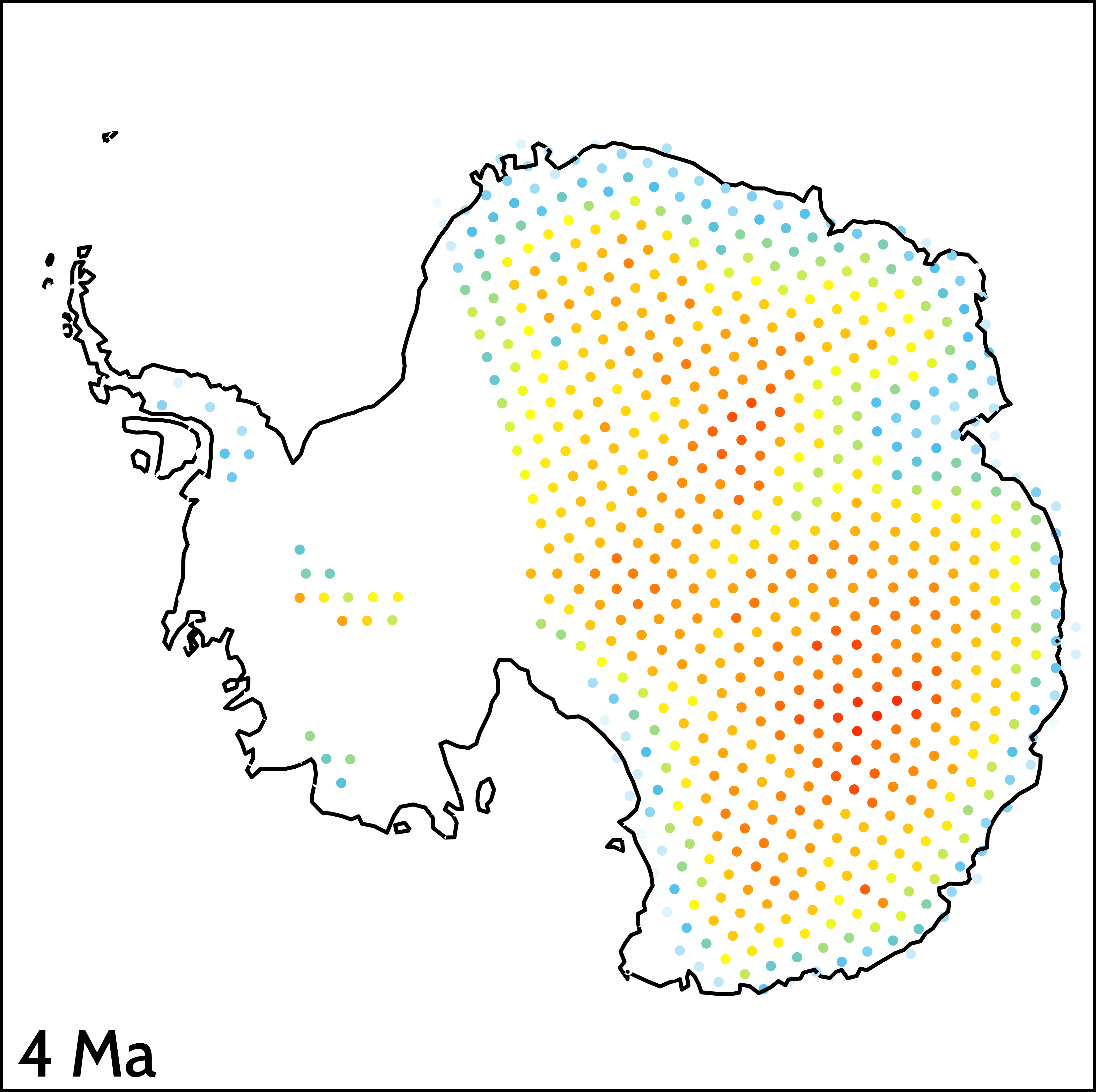}\hfill\\
             \includegraphics[width=0.3\textwidth,height=4.25cm]{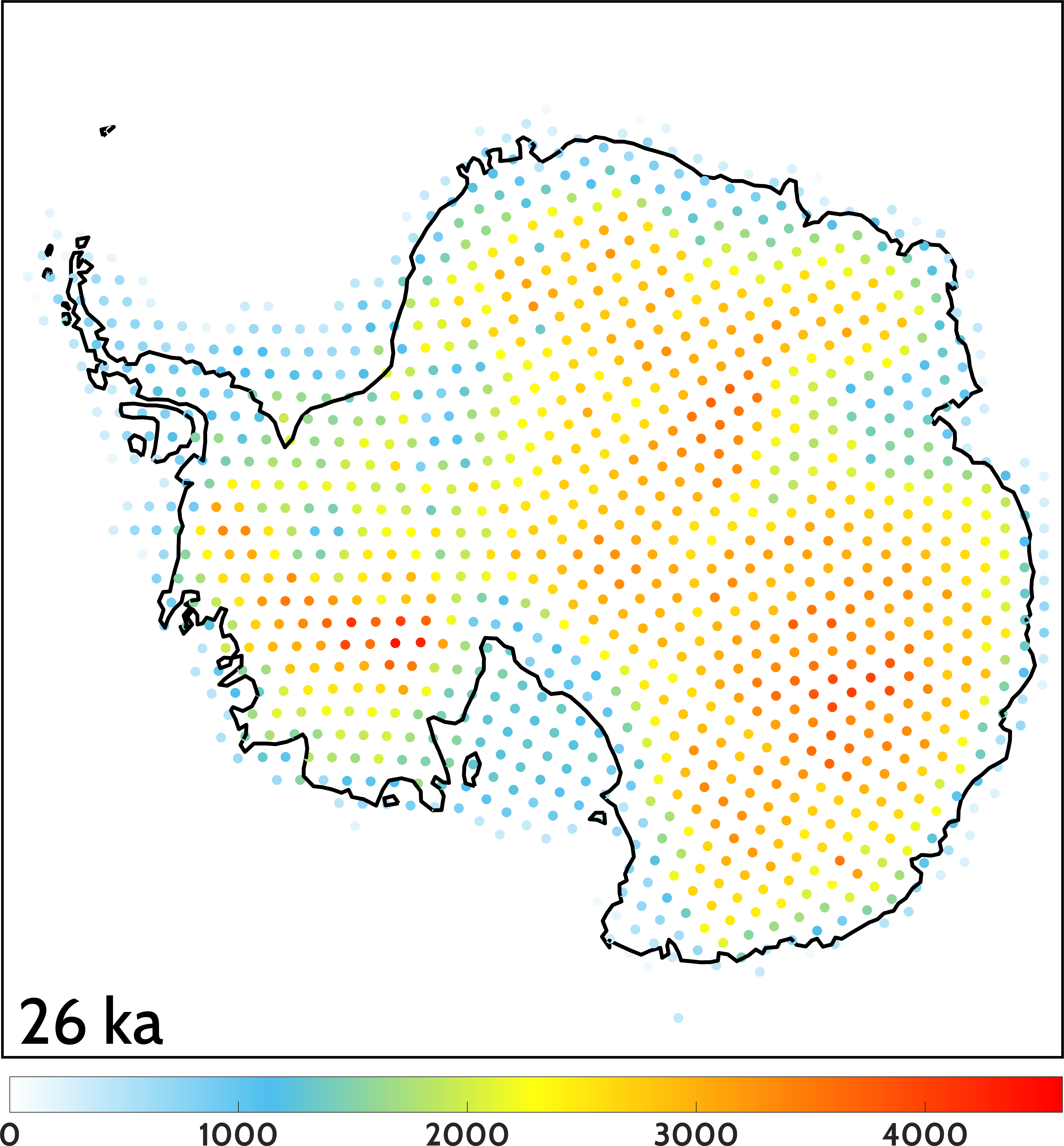}\hfill
    \caption{Time slices of the  Antarctic ice distribution on the pixelated surface of the Earth. Color coding represents the ice thickness in meters. Each spatial spread in used over a specific interval of time (see text).  }
    \label{AIS_Distribution}
\end{figure}
%\begin{figure}[ht!]
%\centering
%     \begin{subfigure}[b]{0.3\textwidth}\hfill
%         \includegraphics[width=\columnwidth, height=4.5cm]{figs/AIS_Minimal2.png}\hfill
%    \end{subfigure}  \\
 %     \begin{subfigure}[b]{0.3\textwidth}\hfill
%         \includegraphics[width=\columnwidth, height=4.5cm]{figs/AIS_Oligo.png}\hfill
%    \end{subfigure}  \\
%      \begin{subfigure}[b]{0.3\textwidth}\hfill
%         \includegraphics[width=\columnwidth, height=4.5cm]{figs/AIS_mio_3.png}\hfill
%    \end{subfigure}  \\
%      \begin{subfigure}[b]{0.3\textwidth}\hfill
%         \includegraphics[width=\columnwidth, height=4.5cm]{figs/AIS_early_plio.png}\hfill
%    \end{subfigure}  \\
%    \begin{subfigure}[b]{0.3\textwidth}
%         \includegraphics[width=\columnwidth,height=4.5cm]{figs/AIS_LGM2.png}\hfill
%    \end{subfigure}  
%    \caption{Time slices of the  Antarctic ice distribution on the pixelated surface of the Earth. Color coding represents the ice thickness in meters. Each spatial spread in used over a specific interval of time (see text).  }
%    \label{AIS_Distribution}
%\end{figure}
As discussed in the main text, the glacial surface loading function is discretized spatially in order to facilitate the computation of the surface integrals by the pseudo spectral approach \citep{mitrovica1991postglacial}. The global estimate of ice volume, M20 \citep{miller2020cenozoic}, is thus distributed over the spherically pixelated grid on the surface of the Earth. \texttt{SELEN}$^4$ adopts the grid of the equal-area, icosahedron-shaped pixels \citep{tegmark1996icosahedron}. This grid is characterized by a resolution parameter $R$ such that the total number of pixels on the surface $P$ is 
\begin{equation}
    P=40R(R-1)+12
\end{equation}
All our simulations were performed with $R=30$, yielding $P=34812$. For such a large number of pixels, each pixel can be thought of as a disk of radius $ r_{disk}\approx 2a/\sqrt{P}$. Thus each pixel approximately covers a surface area of $14600$ km$^2$.

\begin{figure}
\centering
     \begin{subfigure}[b]{0.3\textwidth}\hfill
         \includegraphics[width=\columnwidth, height=4.5cm]{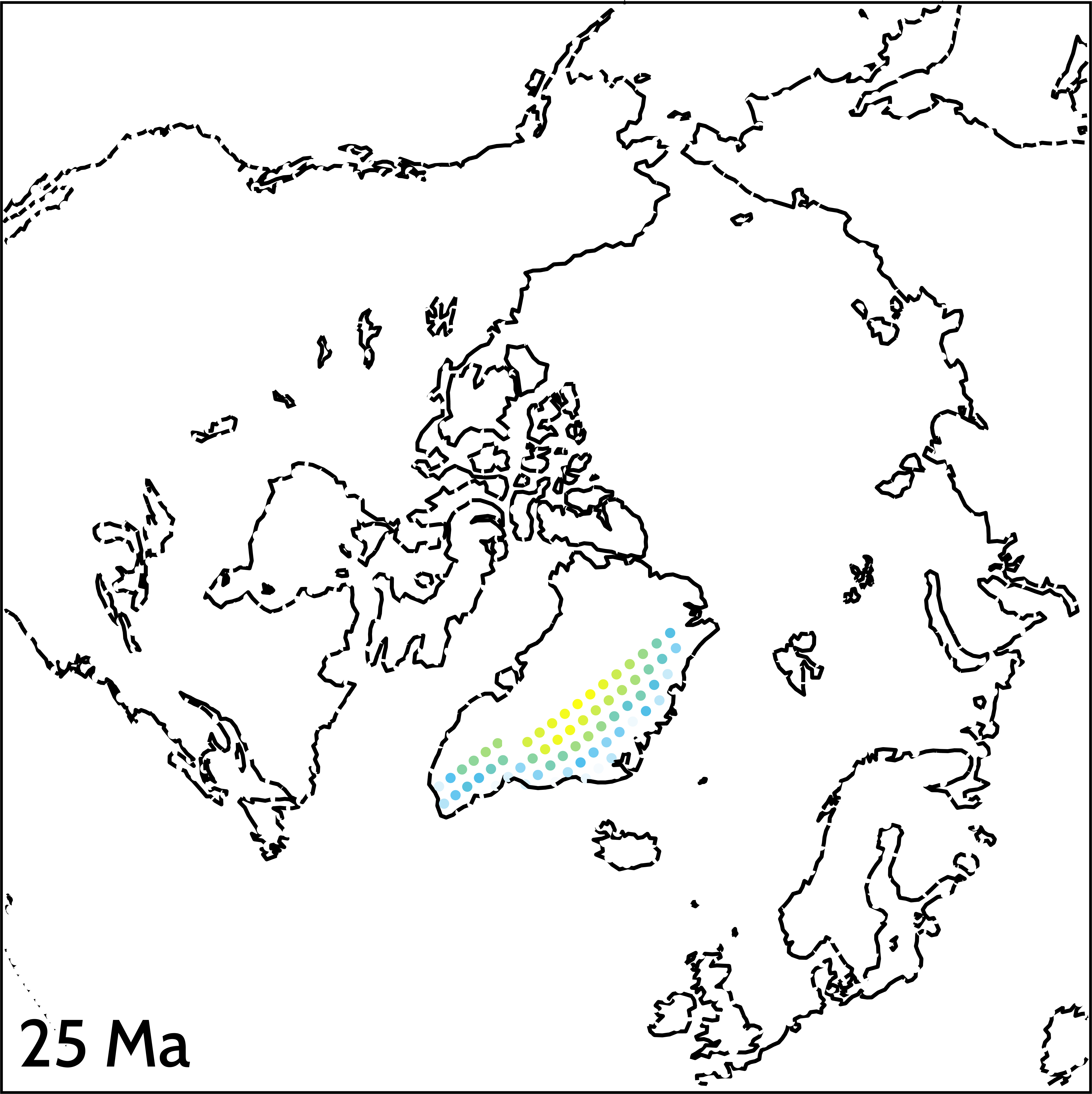}\hfill
    \end{subfigure}  
      \begin{subfigure}[b]{0.3\textwidth}\hfill
         \includegraphics[width=\columnwidth, height=4.5cm]{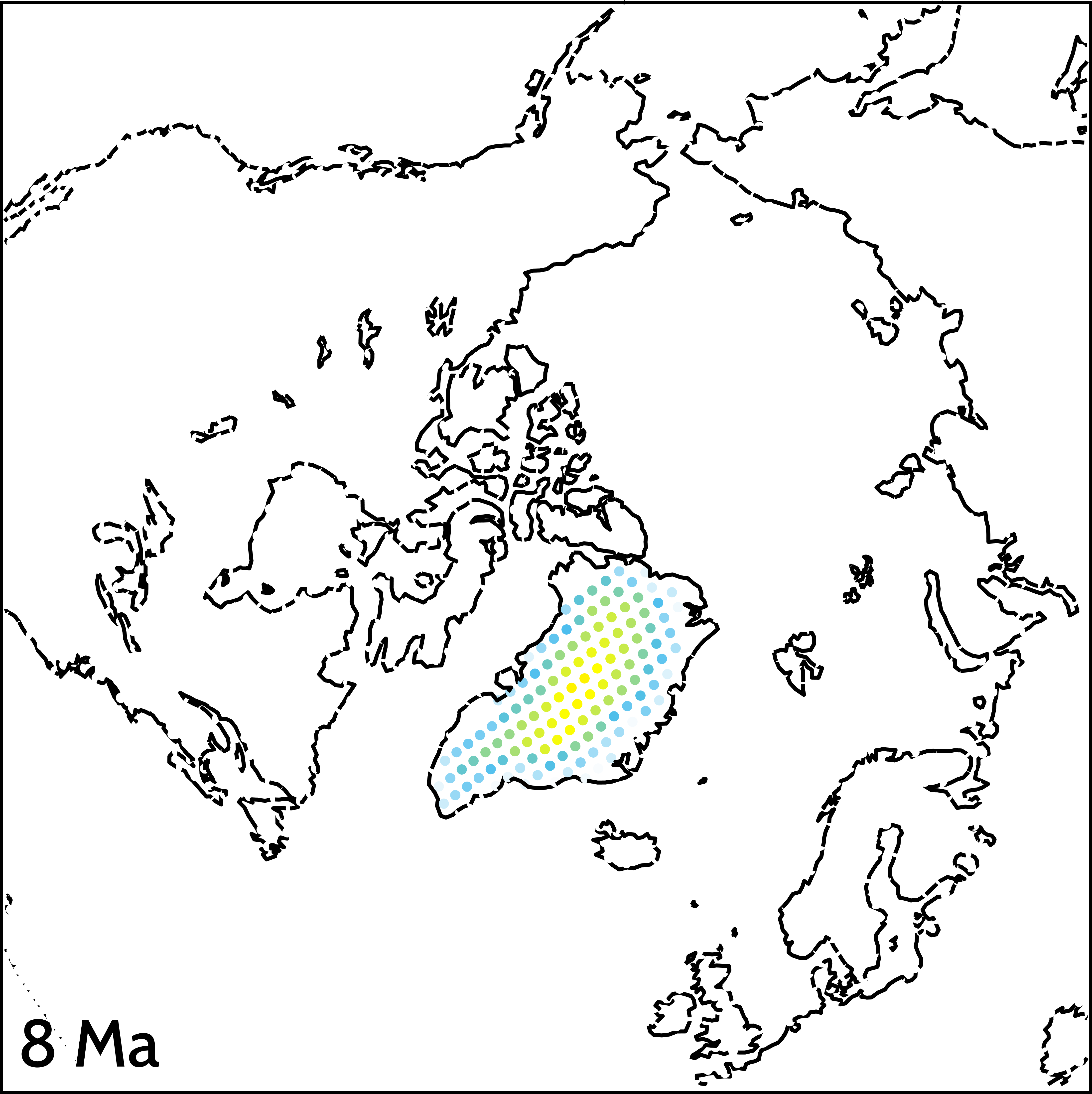}\hfill
    \end{subfigure}  \\
      \begin{subfigure}[b]{0.3\textwidth}\hfill
         \includegraphics[width=\columnwidth, height=4.5cm]{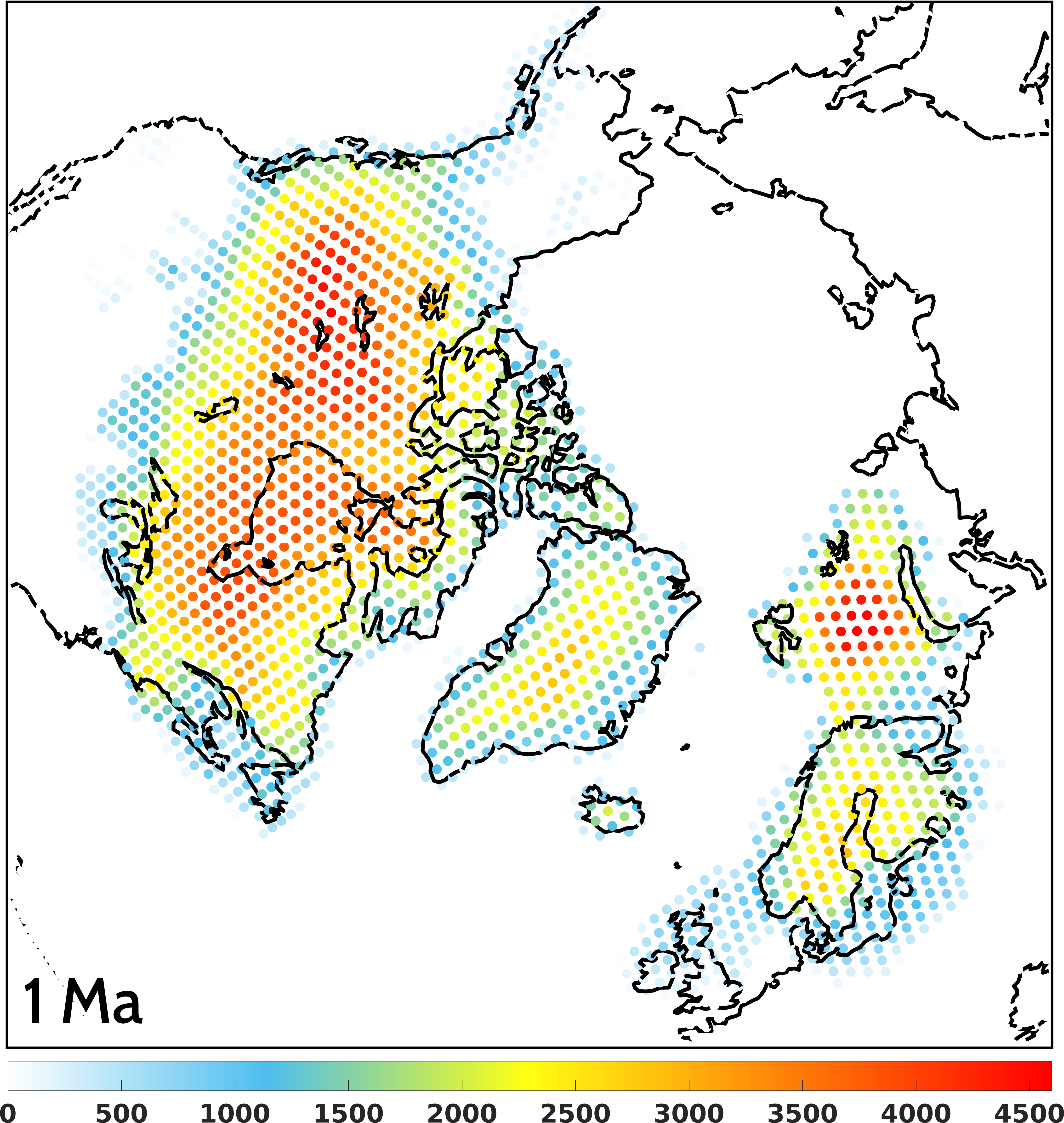}\hfill
    \end{subfigure}  
    \caption{Same as Fig.\ref{AIS_Distribution}, but for the glacial spread over the northern hemisphere.} \label{NH_distribution}
\end{figure}

 In the main text, we identified major milestones of ice variability  based on a compilation of available geologic evidence and GCMs (Table \ref{geologic_refs}). Over the time interval of our simulations (47 Myr), we scale our glacial input with the global limit of M20, however, we separate between 8 intervals of spatial spread. Samples of the spread over these intervals are plotted via \textsc{Matlab}'s Mapping Toolbox \citep{greene2017antarctic} and shown in Fig.\ref{AIS_Distribution} for the Antarctic ice sheet, and in Fig.\ref{NH_distribution} of the northern ice cap. Before the Eocene-Oligocene transition, ice is distributed over the high elevation regions of the East Antarctic Ice Sheet (EAIS); specifically on the Dronning Maud Land, the Gamburtsev Mountain, and parts of the Trans-Antarctic Mountains. The second interval, the Oligocene, witnesses continental scale spread on the EAIS, with minimal glaciation on the high elevation plateaus of the western part (WAIS), and on the Eastern side of the Greenland Ice Sheet (GIS). The third interval, covering the Early Miocene, only differs from the second interval by the expansion of the EAIS into marine terminating glacial spread. The fourth interval represents the warm period of the Middle Miocene Climatic Optimum (MMCO), and is characterized by Antarctic glacial retreat reaching the spread of the first interval. Following the MMCO, the stable EAIS is established, and ice fully covers the WAIS and the GIS for the first time. The sixth interval represents the warm period of the Early Pliocene, and it witnessed the retreat of the WAIS. Starting 3 Ma, the seventh interval highlights the maximum glacial spread over the Cenozoic, with the glaciation of the Laurentide and the Fennosacida ice sheets (Fig.\ref{NH_distribution}). In this interval, for the separation of the global ice volume between the northern and southern regions, we use the simulation of \cite{pollard2009modelling} as an estimate of Antarctic ice. Subtracting this estimate from the global limit leaves us with the contribution of the northern cap. Finally, we terminate our established history with the ICE-6G model \citep{argus2014antarctica}, and we use the spread of the Last Glacial Maximum (LGM) as the limit of the maximum possible ice spread. In each interval, the distribution of the allocated ice volume over the pixels of the corresponding region is  controlled by the relative distribution among the same pixels at the LGM. 

%Appendix sections are coded under \verb+\appendix+.

%\verb+\printcredits+ command is used after appendix sections to list 
%author credit taxonomy contribution roles tagged using \verb+\credit+ 
%in frontmatter.

%\printcredits

%% Loading bibliography style file
%\bibliographystyle{model1-num-names}
\bibliographystyle{cas-model2-names}

% Loading bibliography database
\bibliography{References}

%\vskip3pt

\end{document}